\documentclass{aa}
\usepackage{graphicx}

\begin{document}
\title{The mass and luminosity functions and the formation rate of DA white
dwarfs in the Sloan Digital Sky Survey}

\author{Qi Hu\inst{1, 2}
\and Chaolun Wu\inst{2}
 \and Xue-Bing Wu\inst{3}} \offprints{Xue-Bing Wu}
\institute{Yuanpei Program, Peking University, Beijing, 100871,
China
 \and School of Physics, Peking
University, Beijing, 100871, China
 \and Department of Astronomy, Peking University, Beijing, 100871,
 China;
wuxb@bac.pku.edu.cn }
\date{}
\authorrunning{Hu, Wu \& Wu}
\titlerunning{DA white dwarfs in the SDSS}

\abstract
{}
 {The SDSS Data Release 1 includes 1833 DA
white dwarfs (WDs) and forms the largest homogeneous sample of WDs.
This sample provides the best opportunity to study the  statistical
properties of WDs.}
 {We adopt a
recently established theoretical model to calculate the mass and distance of
each WD using the observational data. Then we adopt a  bin-correction
method to correct for selection effects and use the $1/V$ weight-factor method
to calculate the
luminosity function, the continuous mass function and the
formation rate of these WDs.}
{The SDSS DA WD sample is incomplete and suffers
seriously from selection effects. After corrections for the selection
effects, only 531 WDs remain. From this final sample we
derive the most up-to-date luminosity function
and mass function, in which we find a broad peak of WD masses centered around
0.58$M_{\odot}$. The DA WD space density is calculated as
$8.81\times10^{-5}pc^{-3}$ and the formation rate is $2.579\times
10^{-13}pc^{-3}yr^{-1}$.}
 {The
statistical properties of the SDSS DA WD sample are generally in
good agreement with previous observational and theoretical
studies, and provide us information on the formation and evolution
of WDs. However, a larger and more complete all-sky WD sample is still needed
to explain some subtle disagreements and unresolved issues.}


\maketitle

\section{Introduction}

Studies of white dwarfs (WDs) have 
developed substantially in the last century. 
Chandrasekhar (1933, 1939) first developed a theoretical WD model by applying
Fermi-Dirac statistics of electron and predicted a
relationship between the mass and radius (M--R relation) of the WD. 
Hamada \& Salpeter (1961) improved the model for the
zero-temperature degenerate configuration by incorporating the
assumption of various cores (H, He, C, O, Si, Mg, Fe) for
different WDs. Wood (1990, 1995)
considered more details such as the finite-temperature
effect on the radius and the envelope of WDs, and derived 
theoretical models by calculating the stellar evolution. These models
were widely employed in the following decade. Two of the
latest model calculations are those of Panei et al (2000) and Fontaine et
al. (2001).
On the observational side, thousands of WDs have been detected in
large sky surveys, such as those of the Extreme Ultraviolet Explorer (EUVE),
 the Palomar--Green Surveys (PG), ROSAT All-sky Survey, 2DF QSO
Survey and the Sloan Digital Sky Survey (SDSS). Various spectral
and photometric parameters  of these WDs have been obtained.
McCook \& Sion (1999) published a catalog including 2245
spectroscopically identified WDs. The SDSS Data Release 1
(DR1) included 2551 identified WDs (Kleinman et al. 2004), 
of which only a few WDs
were already in the catalog of McCook \& Sion (1999). The
Hipparcos data (Schmidt 1996; Provencal et al. 1998) provided a
reliable source of the proper motions and parallaxes of a few
WDs. The orbital parameters of WDs in visual
binaries (e.g. radial velocities of WDs in common proper motion
(CPM) systems) can also be obtained from observations (see e.g.
Thorstensen et al. 1978, Vennes et al. 1999, Wegner \& Reid 1987,
Wegner et al. 1989).

The
effective temperature $T_{eff }$ and surface gravity $log\ g$ can
be derived from fitting the Balmer line profiles of WDs.
 Detailed discussions about the fitting techniques can be
found in Bergeron, Saffer \& Liebert (1992, hereafter BSL) and
references therein. $T_{eff}$ can also be derived from the
photometric colors by using an atmosphere model. To test
the theoretical M--R relation, we need to estimate the mass and 
radius of WDs by directly measuring the flux,
distance and the gravitational redshift of them. The distance of
a nearby WD can be obtained directly from the measured parallax.
The gravitational redshifts of some WDs in the CPM systems have
been measured (see Wegner \& Reid 1987; Wegner et al. 1989).
However, these direct measurements can be done only for a few WDs.

Calculating the mass of WDs from the fundamental parameters
mentioned above (such as  $T_{eff }$ and $log\ g$) is the key to obtain 
the mass distribution of a large WD
sample. Currently there are four kinds of methods to
determinate the WD mass (see section 3 for details). There are a number of
previous determination of WD masses (Koester, Schulz \& Weidemann
1979, Weidemann \& Koester 1984, McMahan 1989, Weidemann
1990, BSL, Finley et al. 1997, Marsh et al. 1997a, b,
Vennes et al. 1997, Vennes 1999, Napiwotzki et al. 1999,
Madej et al. 2004; Liebert et al. 2005).  All of these obtained the
mass distribution based on the spectroscopic WD masses, while BSL,
Bergeron, Liebert \& Fulbright (1995) and Reid (1996) obtained the
mass distribution based on the gravitational redshifts. 
Although the gravitational redshift measurements are certainly
important, these can be obtained only for a few WDs. 
For a large sample of WDs, estimating their masses from
the spectroscopic data is probably the only possible way. 

Generally, the mass distribution, the luminosity and mass
functions (LF and MF) of the WDs can be constructed when the
sample of WDs is large and complete enough.  
The luminosity function (LF) and the mass function (MF) derived from a
sufficiently large sample of WDs in the solar neighborhood are
very helpful for the study of the WD formation history. The LF
reveals the current formation rate or death rate of stars in the
local Galactic disk, and the MF can display the roles of close
binary evolution in the WD formation process
(Schmidt, 1959,
1963, 1968, 1975; Green 1980; Fleming et al. 1986; Liebert
et al. 2005). In particular, Liebert et al. (2005) studied
the mass distribution of a volume-limited sample, and obtained the
luminosity and mass functions and the recent formation rate of DA
WDs based on the 348 hot WDs from the PG survey.

The aim of the present paper is to derive the mass distribution
and luminosity and mass functions of the large sample of DA WDs in
the SDSS DR1, taking  advantage of the
larger volume of SDSS DR1 to obtain more reliable
results. We have investigated the whole 1833 DA
white dwarfs in the SDSS DR1, and calculated the mass, radius,
bolometric magnitude, cooling age, and distance of
these WDs using the recently published theoretical evolutionary models
of Panei et al. (2000). These models cover a broader and denser 
parameter space than the models of Wood (1990, 1995), and are more able to 
be applied to obtain the spectroscopic masses for a large sample of WDs.
Based on the derived parameters, we constructed the
luminosity function, mass function and determined the formation rate of DA
WDs. Due to the magnitude-limited selection effect of SDSS, the
sample is far from the completeness needed to obtain statistically reliable
results. Therefore, after introducing the sample and mass estimation method
in section 2, we investigate the sample completeness and correct the 
selection effects in section 3.  Then 
we study the luminosity function, mass function, formation rate and 
3-dimensional distribution properties of DA WDs in sections 4 to 7. 
We briefly summarize and discuss 
our results in section 8.

\section{The SDSS DA WD sample and the mass estimates with the
evolutionary model}

The SDSS is an ongoing imaging and spectroscopic survey of about
ten thousand square degrees in the north Galactic cap
to determine the brightness, positions, and obtain
optical spectra of various objects (York et al.
2000). Although it mainly focuses on the extragalactic objects,
there are many Galactic spin-off projects of which one is to
acquire high-quality stellar spectra from stars of different
spectral types. The spectroscopic survey in the SDSS DR1 covers a
area of 1360 $deg^2$. Kleinman et al. (2004) published catalogs of
the spectroscopic WD and hot subdwarf sample from the
SDSS DR1 (Abazajian et al. 2003). They presented the spectral
fitting results of 2551 certain WDs, 240 hot subdwarf
stars and another 144 possible, but uncertain WDs  and hot
subdwarf stars. In this paper, we
use the spectral data of 1833 DA WDs. Kleinman
et al. (2004) derived the effective temperatures ranging from
7220K to 93855K, and the surface gravities $\log g$ from 6.25 to
9.00 (in $cgs$ units), using the pure hydrogen
atmosphere model of Koester et al. (2001). The photometric
parameters include the five magnitudes in $ugriz$ system (the
magnitudes at g band being from 15.20 to 20.55), the proper
motion velocity, the extinction index of g band and the value of
signal to noise. The WD data of the SDSS DR1 are available
at the SDSS website \footnote{
http://www.sdss.org/dr1/products/value\_added/wdcat/dr1/}.

Previous studies (e.g. Clemens 1993; Barstow et al. 1993) provided
evidence that a DA white dwarf most likely has a thick
hydrogen layer of about $q(H)=-4$. In addition, the suggestion that
most of high mass DA WDs  have a C/O core
 and the low-mass DA WDs have a helium core has been widely
accepted in former studies (e.g. Napiwotzki et al. 1999, Madej, 
Nalezyty \& Althauset 2004, Liebert et al. 2005).
 Theoretical studies also show that when the mass of a WD is less than
0.45 $M_{\odot}$, the progenitors of these WDs could not reach high
enough central temperatures for helium to be ignited at the
center (Panei et al. 2000), providing further evidence for a helium core 
of low-mass DA WDs.
 BSL, Finley et al. (1997), Marsh et al. (1997a, 1997b),
Vennes et al. (1997), Vennes (1999) and Liebert et al. (2005) all
adopted Wood's evolutionary model (Wood 1990, 1995) of a C/O-core 
with a thick hydrogen envelope in their studies.
Madej et al. (2004) brought in the evolutionary
models of Panei et al. (2000) assuming a C/O core with
$M(H)/M=10^{-5}$ in the atmosphere of massive WDs and a helium
core with $M(H)/M=3\times10^{-4}$ for WDs with a mass less than 0.45
$M_{\odot}$. In this paper, we use the evolutionary helium-core
model for WDs with mass less than 0.45 $M_{\odot}$
and bring in the C/O-core model for those with mass
larger than 0.45 $M_{\odot}$. Both models are from Panei et al. (2000).
In this
paper we
assume that all DA WDs (with either a C/O or helium core) have a thick
hydrogen layer ($q(H)=-4$, $z=0.001$) for simplicity.

\begin{figure}
    \includegraphics[width=9.2cm, height=7.5cm]{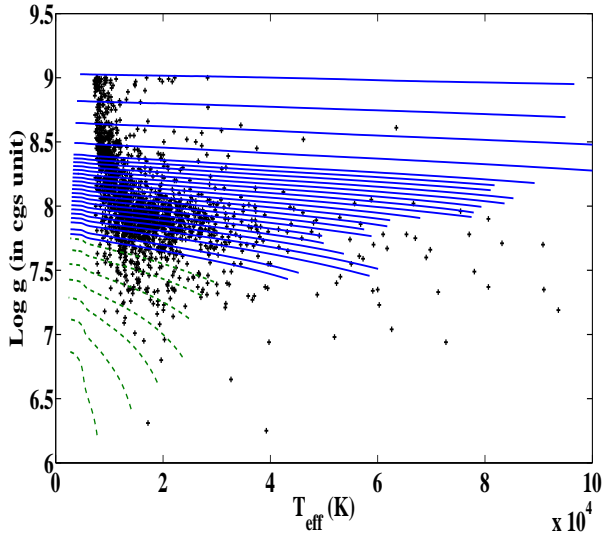}
    \caption
{ The $\log g$ - $T_{eff}$ diagram. The lines represent the theoretical
results from the evolutionary model of Panei et al. (2000). The seven dashed
 lines at the bottom represent the results for the He-core WDs
with a hydrogen layer. The solid lines represent the results for
the C/O-core WDs with a hydrogen layer. These lines correspond to masses of
(from bottom to top)
0.15, 0.2, 0.25, 0.3, 0.35, 0.4, 0.45 $M_{\odot}$ for He-core WDs, and
0.45, 0.47, 0.50, 0.52, 0.54, 0.56, 0.58, 0.60, 0.62, 0.64, 0.68,
0.70, 0.72, 0.74, 0.76, 0.78, 0.80, 0.82, 0.84, 0.90, 1.0, 1.10,
1.20 $M_{\odot}$ for C/O-core WDs respectively. The 1833 DA WDs of SDSS DR1
are plotted as crosses.} \label{Fig.1}
\end{figure}

The predictions of the models of Panei et al. (2000)
and the data ($\log g$, $ T_{eff}$) of all DA non-magnetic WDs 
in the SDSS DR1 are
plotted in Fig. 1. From it we can see that the models we chose
are appropriate for our study. The predicted parameters of two models
(He-core and C/O-core) cover a
lager parameter range in the figure. The C/O models cover 
$T_{eff}$ from about 4000K to 100000K, and  $log\ g$ from 7.43
to 9.03, while the He-core models cover $T_{eff}$ from
2500K to 27000K and $log\ g$ from 6.2 to 7.7. Most observed data
of the DA WDs in the SDSS DR1 are within the range of these models.
Some high temperature WDs at the right part of Fig. 1
are not covered by any models. The parameters of
these extremely high-temperature WDs were discussed in BSL (
they limited the $T_{eff}$ of their WD sample to less than 40000K).
There is a large discrepancy between the
effective temperatures obtained from two different methods of
fitting the Balmer lines  when the $T_{eff}$ of a WD is above
50000K. We eliminate the WDs with extremely high $T_{eff}$
($T_{eff}$ $>$ 48000K) from our samples. Because these WDs
( 39 objects) are only a very small fraction (about 2\%) of the
whole sample, their influence on the
completeness of the sample is small.

To ensure that our calculation of WD parameters through the evolutionary
model is
reliable,  we tested the model by comparing the WD masses derived
 from the evolutionary model and spectroscopic parameters
($T_{eff}$ and $\log g$) with the masses obtained from other independent
methods. Currently there are three methods to
determine the masses of WDs  without involving the evolutionary
models, namely:

(1) If a WD is in a binary system, we can precisely calculate the
WD mass from its orbital parameters, but systems with
complete spectroscopic parameters are relatively rare. Only a few
WDs, including Sirius B (Gatewood \& Gatewood 1978;  BSL; Provencal et
al. 1998; Barstow et al. 2005), 40 Eri B (Shipman et al. 1997; Wegner 
1980; Finley et
al. 1997), and Feige 24 (Vennes et al. 1991), have two independent
estimates of their masses.

(2) If a WD does not have the orbital data, but has a parallax value
from which the distance of it can be derived, we can obtain its
absolute magnitude $M_V$ from  $M_V  = V + 5
+ \log \pi$, where $V$ is the V band magnitude of the star 
 and $\pi$ is the parallax in units of $arcsec$. Then we can
obtain the bolometric magnitude from $M_{bol}= M_V  + B.C._V$, where
$B.C.$ is the bolometric correction in the $V$ band, which can be
estimated by the interpolation of the grid of the atmosphere model
(Bergeron et al. 1995b).
Then
from equations $M_{bol}  =  - 2.5\log (L/L_\odot) +
M_{bol, \odot }$ and $ L = 4\pi R^2 \sigma T_{eff}^4$ 
(here $\sigma$ is the Stephan-Boltzmann constant), the radius $R$
can be derived.
By using Newton's law of gravitation, we can obtain the mass of a
WD.

(3) In some cases, the WD mass can be derived from the  gravitational
redshift $\Delta \lambda /\lambda$, which is usually described by
the equivalent Doppler shift velocity: $v = c\frac{{\Delta \lambda
}}{\lambda } = 0.6362\frac{M}{{M_ \odot }} \cdot \frac{{R_ \odot
}}{R}(km/s)$. So if we know the radius (or mass) from the second
method mentioned above, we can easily derive the  mass
(or radius) accordingly.

In Table 1 we list the parameters and the references of the WDs
which have both spectroscopic mass (derived from $T_{eff}$ and
$\log g$ by using the evolutionary model of Panei et al. (2000))
and mass derived from one of the other three methods without using
the evolutionary model. We note that BSL, Bergeron et al. (1995a),
Provencal et al. (1998), and Boudreault \& Bergeron (2005) have
compared the masses of some WDs derived from different
methods. Here we incorporate these WDs in Table 1
and re-estimate the spectroscopic masses of WDs
using the new evolutionary model of Panei et al. (2000).
We also add dozens of DA WDs that are not included in these
previous studies.

\begin{figure}
    \begin{center}
    \includegraphics[width=9.2cm, height=7.5cm]{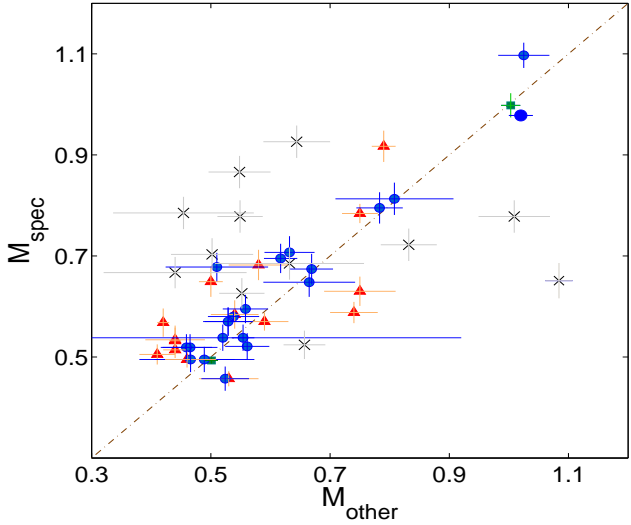}
    \caption{A comparison between the WD mass derived from the evolutionary
model and those determined by other methods without using a theoretical
M-R relation. Both axes are in unit of $M_{\odot}$. All points in the diagram
are listed in Table 1. Triangles represent the WDs with mass determined by
the triangle parallaxes and surface gravity. Circles represent WDs with
 mass determined by the gravitational redshifts. Squares represent
WDs with mass derived from the orbital parameters. Crosses represents WDs
 cooler than 12000K. The dashed line represents $M_{spec}=M_{other}$.}
    \label{Fig.2}
\end{center}
\end{figure}

Fig. 2 shows the comparisons of the WD masses obtained from different methods.
WDs were divided into two groups, one with $T_{eff}$ less than 12000K
(represented by crosses) and the other with $T_{eff}$ higher than
12000K. We can see that except for several
WDs with
$T_{eff}$ less than 12000K the masses estimated with different methods
 are in good agreement.
For WDs with $T_{eff}$ less
than 12000K, the differences in masses estimated by different methods
are obviously larger. This is because these cooler WDs
are likely to be convective. BSL have convincingly proved that
the convection effect leads to significant amounts of helium (which is
invisible in the spectra) entering the atmosphere, 
producing higher pressure which would substantially affect the spectral
line profiles. The total effect on the spectral line is
indistinguishable from the increased surface gravity. In other
words, a low-temperature DA WD with large surface gravity might
actually be a helium-rich star with lower surface gravity (and correspondingly
with lower mass). So the scatter in the masses
estimated with different methods for cooler WDs possibly has less
to do with the evolutionary model that we adopted but is mainly due to
the techniques of analyzing the spectral lines. For this reason, we
remove these WDs from our statistical analyses.

There are still some high-temperature
WDs for which the different mass estimates do not match very well. A
few factors can contribute to this discrepancy, such as the
techniques of fitting the spectral lines, the uncertainties of the
observational parameters. etc. One of the most important
factors is that there seems to be no appropriate evolutionary model for
these high-temperature WDs. For example, G238-44, GD140, EG50, and
EG21 have relatively higher spectroscopic mass compared with the
mass derived from other methods. If we apply a thin hydrogen layer
model ($q(H)=-1$, $q(He)=-4$) or a metal core (like $Fe$ core) for these
four WDs, their spectroscopic mass will be lower by 0.04$\sim$0.06
$M_{\odot}$, and thus the mass comparison of these four WDs would
be better. Moreover, the presence of helium in the atmosphere
would also significantly influence the mass estimate. Boudreault \&
Bergeron (2005) gave a detailed discussion of this effect. They
calculated the masses by using the models of Fontaine et al.
(2001) and assuming a mixed composition in the atmosphere with
$M(He)/M(H)=1$ rather than a pure hydrogen atmosphere, and
obtained similar results that the mean $M_{spec}$ of WDs in their
sample will be lower by 0.2 $M_{\odot}$. Thus, if we adjust
the thickness of the envelope, the composition of the atmosphere
and the atom in the core, more than half of the WDs in Fig. 2 will
have their $M_{spec}$  equal to the mass derived by the
other method. Therefore, we may find the most appropriate
evolutionary model for each WD by matching two kinds of mass
estimates, and then the discrepancy in Fig. 2 would be alleviated.

However, for most DA WDs from SDSS DR1 in our sample,
we do not have parallax or gravitational redshift data to
derive a second mass estimate and do not have further information
about their internal structure and atmospheric composition. 
So we will just assume a theoretically
appropriate model for our samples. From Fig. 2, we find that the
comparison results are satisfactory in general, ignoring the
low-temperature WDs. We then conclude that the assumptions
of evolutionary models we adopted are generally reliable.

After testing the applicability of the model of Panei et al. (2000), 
we use it to calculate the masses of SDSS DA WDs in our sample.
From the $T_{eff}$-$\log g$ diagram shown in Fig. 1,  we can see that using
the two parameters $T_{eff}$ and $\log g$ we can determine the mass of
the WDs (the theoretical lines can be interpolated 
to cover the area that the lines do not cover in the figure). The other
parameters of the WDs also can be calculated based on the mass estimation.

Using the methods described above, we can calculate
 the radius $R$, luminosity $L$ and  bolometric magnitude $M_{bol}$ for WDs.
Similar to the determination of mass, the cooling age of a WD
can be derived by interpolating the grids of the evolutionary
models, thus $\log Age = \log Age(T_{eff},~\log g)$. The bolometric
correction ($B.C.$) in the g band can be obtained by using the model
atmospheres of Bergeron et al. (1995b). $B.C.$ is derived through
interpolating the $T_{eff}$ and $\log g$ into the  grid
of the model atmosphere in Bergeron et al. (1995b) 
for the $ugriz$ system,  $B.C. =
B.C._{Bergeron} (T_{eff} ,\log g)$. The distance $r$ (in $pc$) of the
star can be derived from the $B.C.$ and the relationship between the
absolute magnitude and visual magnitude in the g band: $M_{bol}  = M_g
+ B.C._g $, $M_g  = g - 5 - 5\log {\rm{r}} - {\rm A}_g $, where
$A_g$ is the extinction in the g band which is provided by the SDSS.

To compare our results with other previous work, we also derived
the absolute magnitude of WDs in the V band which were commonly used in
previous studies. Using the results of Bergeron et al.
(1995b), we can easily convert the $M_g$ of the $ugriz$ system to the
$M_V$ of the $UBV$ system. Bergeron et al. (1995b) provide the grids
in both $ugriz$ and $UBV$ systems and the relationship between them,
so we can obtain the bolometric correction in the $V$ band and $M_V$ of
the $UBV$ system by interpolating the $T_{eff}$ and $\log g$ in the
grid. Both $M_V$ and $B.C._V$ (bolometric correction at V band)
are a function of $T_{eff}$ and $log\ g$, namely, $B.C._V  =
B.C._{V,Bergeron} (T_{eff} ,\log g) $, and $M_V  = M_{V,Bergeron}
(T_{eff} ,\log g)$.
We also calculate the galactic coordinates $(l, b)$
of WDs through the equatorial coordinates ($\alpha$, $\delta$) provided
by the SDSS.

The uncertainties in the values of parameters can be estimated in the 
following way. Assuming that a function is determined by several input
parameters, $ f = f(x_1 ,x_2 ,x_3 , \ldots ) $,  the error in this
function $f$  can be expressed as:
\begin{equation}
\delta f = \sqrt {\sum\limits_i^{} {(\frac{{\partial f}}{{\partial x_i }}}
\cdot \delta x_i )^2 } .
\end{equation}
Since all parameters calculated in our study are determined by
$T_{eff}$ and $\log g$, the errors of these parameters are derived
from the errors of $\delta T_{eff}$ and $\delta \log g$, which
have been listed in the catalogue of DA WDs of SDSS DR1 (Kleinman et al.
2004).

In Table 2 we list our main results. As there are 
many WDs in the sample and every WD has many parameters, a subset of
Table 2 only is shown here.
A full table will be provided electronically upon request.

\section{Sample completeness and selection effect correction}

Madej et al. (2004) derived the SDSS WD mass distribution
by counting the number of WDs in the sample. However, the SDSS
detects WDs in the $g$ magnitude range from about 16 to 20 mag, much
fainter than the previous  Palomar-Green Survey
(Fleming et al. 1986; Liebert et al. 2005) and EUVE Survey (e.g.,
Vennes et al. 1997; Finley et al. 1997; Napiwotzki et al. 1999).
Many WDs are several hundreds or even thousands of $pc$  away from the
Earth, thus the magnitude-limiting selection effect plays an
essential role and the sample is often far from complete. One should
first test the completeness of the sample and make necessary
corrections to remove the selection effect; otherwise, the
result will be seriously biased.

\begin{figure}
    \begin{center}
    \includegraphics[width=9.2cm, height=7.5cm]{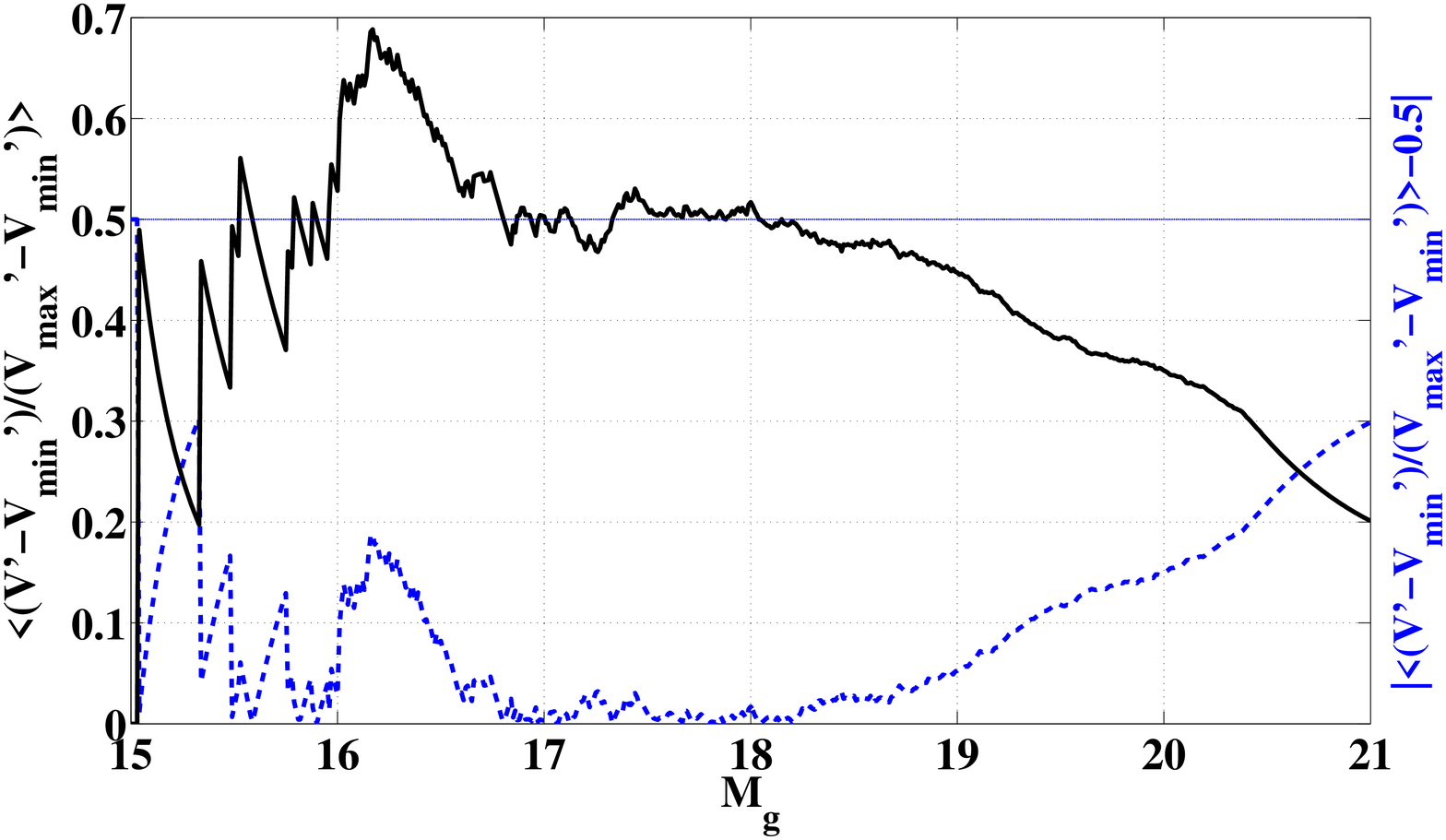}
    \caption{The value of $(V-V_{min})/ (V_{max}-V_{min})$ varies along with
$m_{max}$ (here denoted as $m_g$). In a range from 18 to 21 mag, the trend of
the sample forge toward completeness (i.e. $(V-V_{min})/
(V_{max}-V{min})=0.5$) as the $m_{max}$ decline is obvious.
The sample is of completeness in the range between 17 and 18 mag,
and the majority of WDs in the whole sample stand in this range.
The large fluctuation between 15 and 17 mag is due to less WDs
remaining in this range, which usually leads to large fluctuation in
statistics.} \label{Fig.3}
\end{center}
\end{figure}

We use the method of Schmidt (1968), Green (1980) and Fleming et
al. (1986) to calculate the corrections and derive the WD
Luminosity Function (LF). Generally, for an all-sky WD
sample with upper visual magnitude limit $m_{lim}$, e.g. for $V$
band, given a specific WD with its absolute V magnitude $M_V$ and
distance $r$, the $r$ defines a volume V, and and the $m_{lim}$ define
a maximum distance $r_{max}$ and consequently the maximum volume
$V_{max}$ with the following equation: $M_V = -5\log r_{max}+ m_{lim}+5$
(we assume an average extinction which has been included in
$m_{lim}$). The SDSS WD space distribution scale is about 1 $kpc$,
whereas the galaxy disk radius is about 10 $kpc$, so it is natural
to assume that the WD space radial distribution around the sun
is uniform. To correct the non-uniform height distribution, we
define $dV(z)=exp(-z/z_0)dV$ and adopt $z_0=250pc$ as the scale
height, as done by Fleming et al. (1986) and Liebert et al.
(2005). Thus, if the sample is complete, the average value
of $V/V_{max}$ will be equal to 0.5 (Green 1980). Otherwise, to make
the sample uniform, one should lower the $m_{lim}$ and eliminate
the WDs with $V>V_{max}$ until $\langle V/ V_{max} \rangle=0.5$.

\begin{figure}
    \begin{center}
    \includegraphics[width=9.2cm, height=7.5cm]{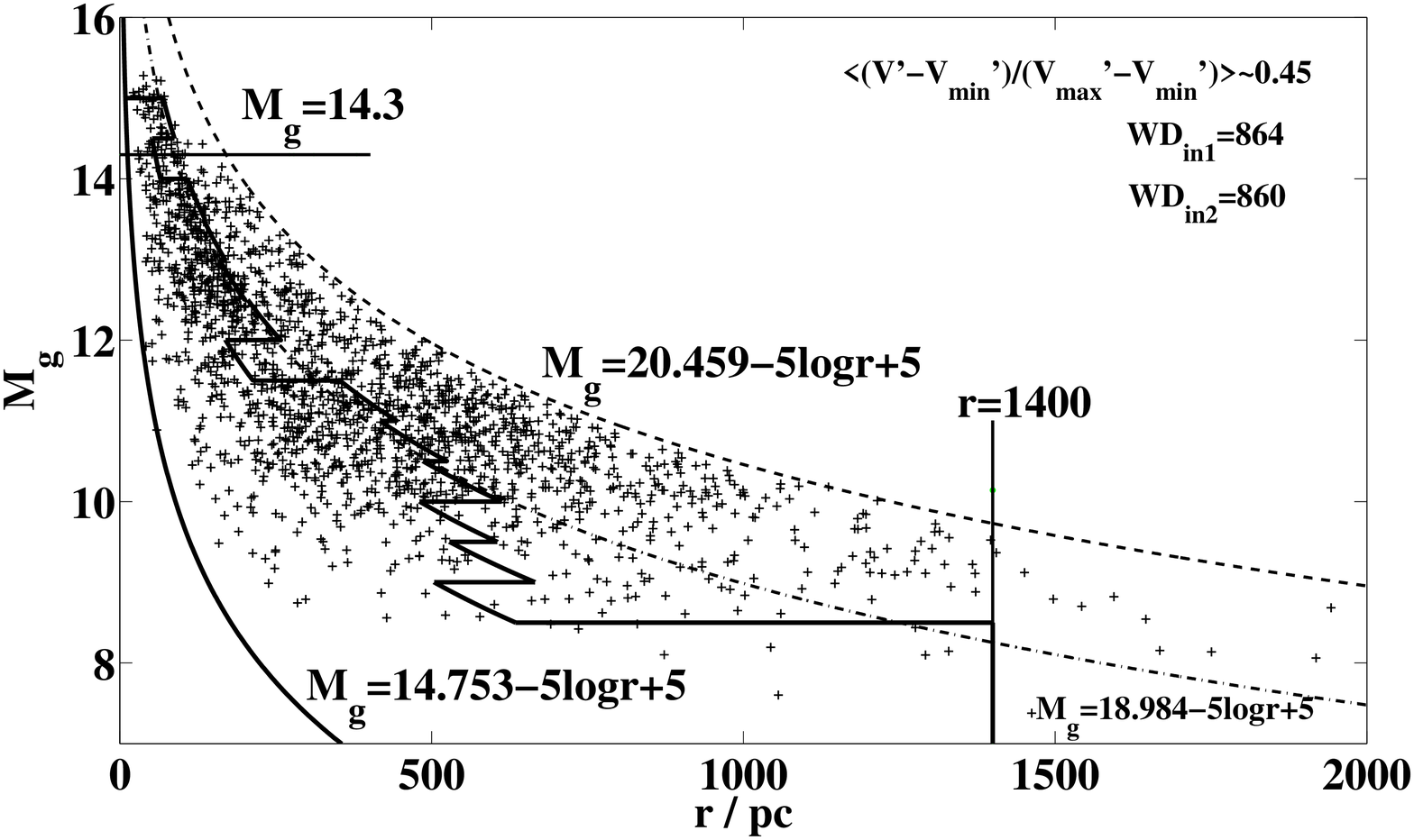}
    \caption{
Selection effect correction according to $g$ band and the
comparison of the two correction methods. The crosses denote the
 1794 SDSS DA WDs with $T_{eff}<48000K$. The left solid line shows
the lower $g$ band magnitude limit 14.753 mag and the right dashed
line shows the upper  magnitude limit 20.459 mag. The middle dot-dashed
line is the uniform-correction result with the upper limit 18.984
mag. The sawtooth curve is the bin-correction result giving the
upper correction boundary $M_g$($r_{max}$). Two straight lines
$Mg=14.3$ and $r=14000 pc$ also give boundaries out of which the
selection effect is relatively severe. There are in total 864 WDs
remaining in the sample after the uniform-correction and 860 after
the bin-correction.} \label{Fig.4}
\end{center}
\end{figure}

Moreover, we make two small changes to the method:

(1) In addition to the upper magnitude limit $m_{max}$, the SDSS
also has a lower magnitude limit $m_{min}$, which defines a
minimum distance $r_{min}$ and a volume $V_{min}$. The reason is
that the SDSS's main focus is extragalactic objects whose
magnitudes are usually faint and the WDs are just its spin-off
projects (Kleinman et al. 2004). Thus the actual space where WDs
were detected is between $r_{min}$ and $r_{max}$, and its volume
is $V_{max}-V_{min}$. So we should use
$\langle(V-V_{min})/(V_{max}-V_{min})\rangle$ instead of
$V_{max}$ to test the sample's completeness.

(2) The SDSS DR1 spectroscopic data cover only 1360 $deg^2$ of the
whole sky (Abazajian et al. 2003), so the
volume is not a spherical or elliptical shape but a cone shape. We
also assume that the $d\Omega(b)$ of the volume $dV$ is apart from
its galactic latitude $b$. So the cubic angle $\Omega$ is simply
1360 $deg^2$. Then we derive the expression of volume $V$ as a
function of $r$ and $b$.
$$V(r,\theta ) = \\
 \int_\Omega  {d\Omega \int_0^r {e^{ - z/z_0 } r^2 dr} } = \frac{{1360}}{{360^2 /\pi }} \cdot 4\pi  \cdot
~~~~~~~~~~~~~~~~~~~~~~~~ \\
  $$
\begin{equation}~~~ (\frac{{z_0 }}{{\cos \theta }})^3 \{ 2 - [(\frac{{r\cos \theta }}{{z_0 }})^2  + 2\frac{{r\cos \theta }}{{z_0 }} + 2] \cdot e^{ - r\cos \theta /z_0 } \},  \\
\end{equation}
where $\theta  = (\pi /2) - b $.

\begin{figure}
    \begin{center}
    \includegraphics[width=9.2cm, height=7.5cm]{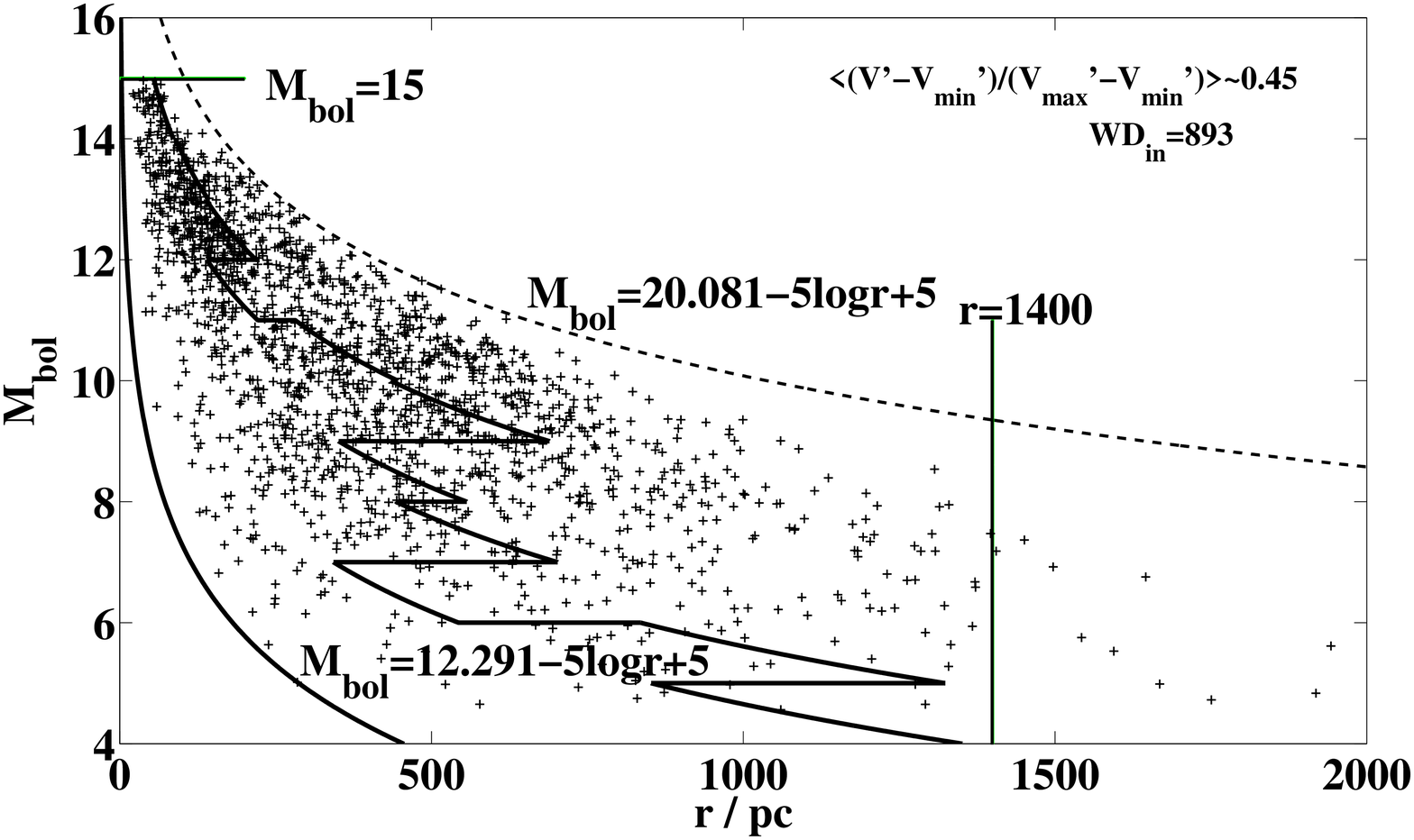}
    \caption{Selection effect bin-correction according to $M_{bol}$.
     The crosses denote the SDSS 1794 DA WDs with $T_{eff}<48000K$.
     The left solid line shows the lower bolometric magnitude limit
     12.291 mag and the right dashed line shows the upper limit 20.081 mag.
     The middle sawtooth curve is the bin-correction result giving
     the upper correction boundary $M_{bol}$($r_{max}$). Two straight
     lines $M_{bol}=15$ and $r=14000$ pc also give boundaries out of
     which the selection effect is relatively severe. There are total 893
     WDs remaining in the sample after bin-correction.}
\label{Fig.5}
\end{center}
\end{figure}

For $g$ band magnitude, the $(V-V_{min})/(V_{max}-V_{min})$ value
of the 1794 non-magnetic SDSS DA WDs with $T_{eff}$
$<$48000K is around 0.3, which suggests that the sample is far
from complete. If we lower the the upper magnitude limit
$m_{max}$(see Fig. 3) to about 18.2 mag,
the $\langle(V-V_{min})/(V_{max}-V_{min})\rangle$ approaches 0.5
and the remaining sample is more complete. However, too many WDs
will be eliminated. We made a compromise: if $\mid
\langle(V-V_{min})/(V_{max}-V_{min})\rangle-0.5\mid <0.05$, the
sample will be regarded as complete. This means that we eliminate the
most selection effect contaminated part of the sample. Setting the
upper limit $m_{max}=18.984$ for $g$ band, the number of WDs in
the remaining sample is 864 out of 1794, almost  half (see Fig.
4, we also required $M_g<14.3$, for the fainter WDs are difficult
to  detect). Similarly, Green (1980) also set $ V'/V_m'
=0.46\pm0.03$ in his study.

An improved method (hereafter named bin-correction, while the
above method is named uniform-correction or
ordinary-correction) is to consider the different magnitude upper limit
for different absolute magnitudes of a specific WD. We
divide the whole sample into 0.5-mag-width bins according to $M_g$
(or 1-mag-width bins according to $M_{bol}$), assuming that the
WDs within the same bins have the same  upper magnitude limit and
the whole sample shares a lower magnitude  limit. In each bin,
$\mid (V-V_{min})/(V_{max}-V_{min})-0.5\mid <0.05$. Fig. 4 
shows this bin-correction for $g$ band. The number of WDs in the
remaining sample is 860. Although the number is more or less the
same as the 864 of the uniform-correction, the difference is obvious:
the  upper magnitude limit at the brighter end is usually smaller than
the 18.984 mag of the uniform-correction. Such WDs observed
are relatively distant, whereas the fainter end's upper magnitude
limit is usually larger than the 18.984 mag, because only the
faint WDs that are near us can be observed and the magnitude-limiting
selection effect is relatively small for the nearby star sample.
Thus we prefer to use the improved bin-correction method.

\begin{figure}
    \begin{center}
    \includegraphics[width=9.2cm, height=7.5cm]{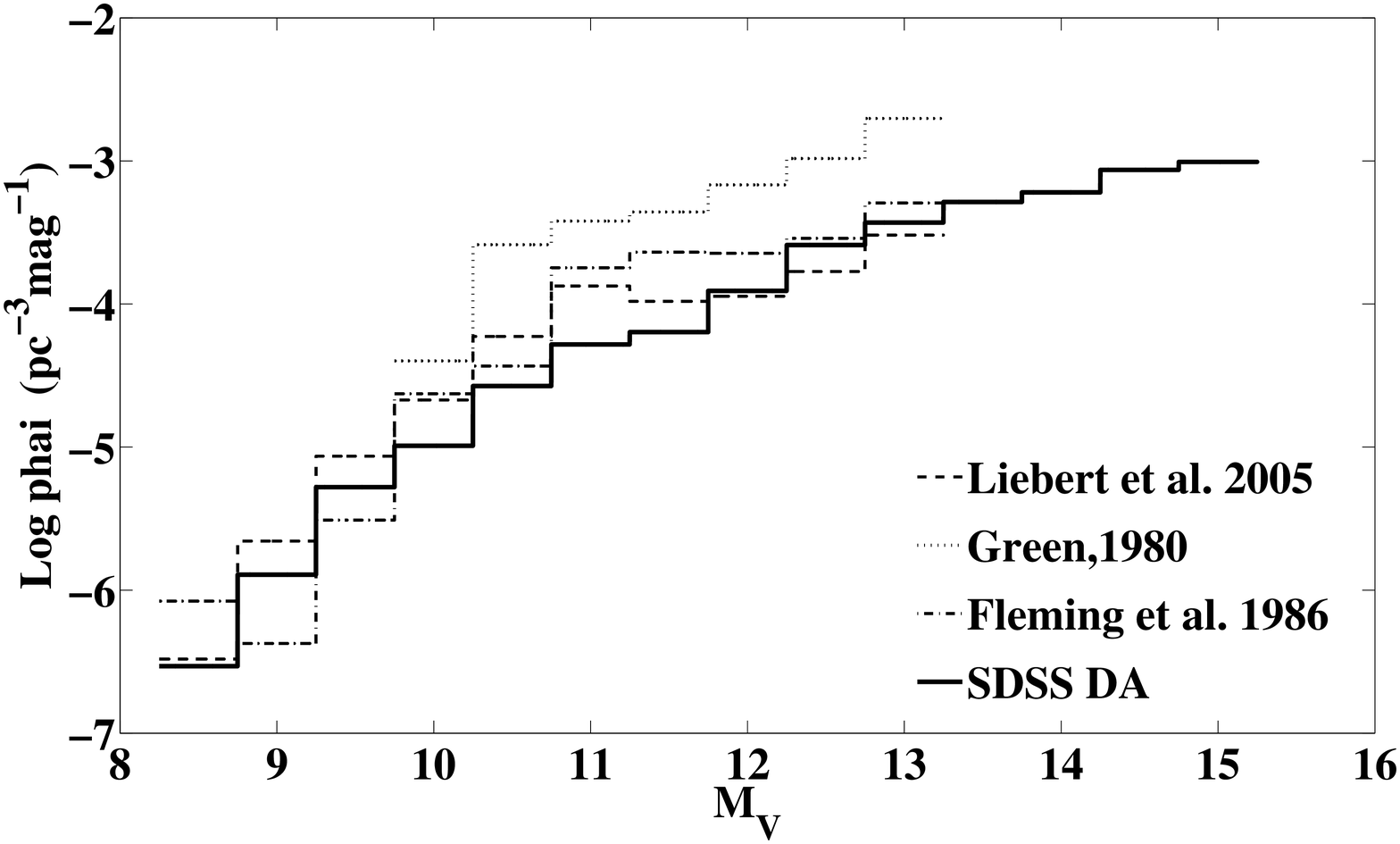}
    \caption{SDSS WD LF as a function of $M_V$, compared with previous LFs:
dashed, dot-dashed and dotted lines denotes the PG LF of Liebert et al.
(2005), LF of Fleming et al. (1986) and LF of Green (1980), respectively.
 Solid lines denote the SDSS DA WD LF.}
\label{Fig.6}
\end{center}
\end{figure}

We also made a 1-mag-width bin-correction
according to $M_{bol}$ for the whole 1794 non-magnetic SDSS DA WD
sample with $T_{eff}<48000$K. The number of WDs in the remaining
sample is 893, as shown in Fig. 5. The reason why we choose $M_{bol}$
as a criterion is that the u, g, r, i, z or V bands have their own
magnitude limits and selection effects. The
extinction is also different from short wavelengths to long
wavelengths. But above all, the $M_{bol}$ can represent all these
factors. After
this selection effect correction, our analysis of SDSS WD samples
will be much less biased and more reliable.

\section{Luminosity function of SDSS DA WDs}

The SDSS WD LF is calculated using the method of Green (1980) and
Fleming et al. (1986). Its distribution volume is
$V_{max}-V_{min}$, and the weight factor is $1/(V_{max}-V_{min})$
which is similar to the $1/V'_m$ used in some previous studies.

\begin{figure}
    \begin{center}
    \includegraphics[width=9.2cm, height=7.5cm]{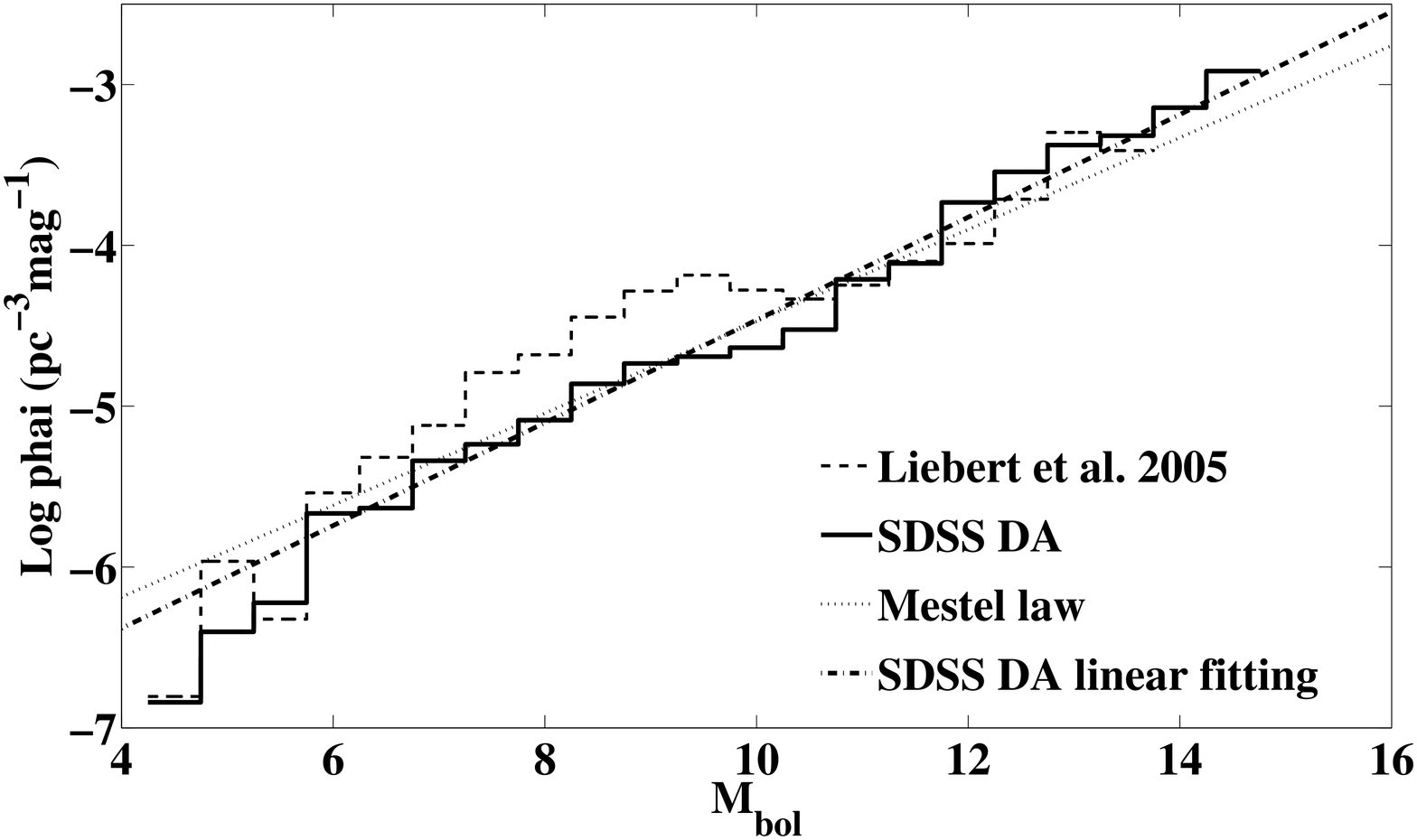}
    \caption{The LF of SDSS WDs as a function of $M_{bol}$, compared with
PG LF (Liebert et al. 2005) and theoretical Mestel (1952) law. The dashed
line denotes the PG LF by Liebert et al. (2005), while the solid
line denotes the SDSS DA WD LF. The dotted-dashed line is the
SDSS DA WD LF linear fitting line $\log \Phi=0.3195M_{bol}-7.660$
and the dotted line represents Mestel's cooling law, $\log
\Phi=0.2857M_{bol}-7.330$.} \label{Fig.7}
\end{center}
\end{figure}

\begin{figure}
    \begin{center}
    \includegraphics[width=9.2cm, height=7.5cm]{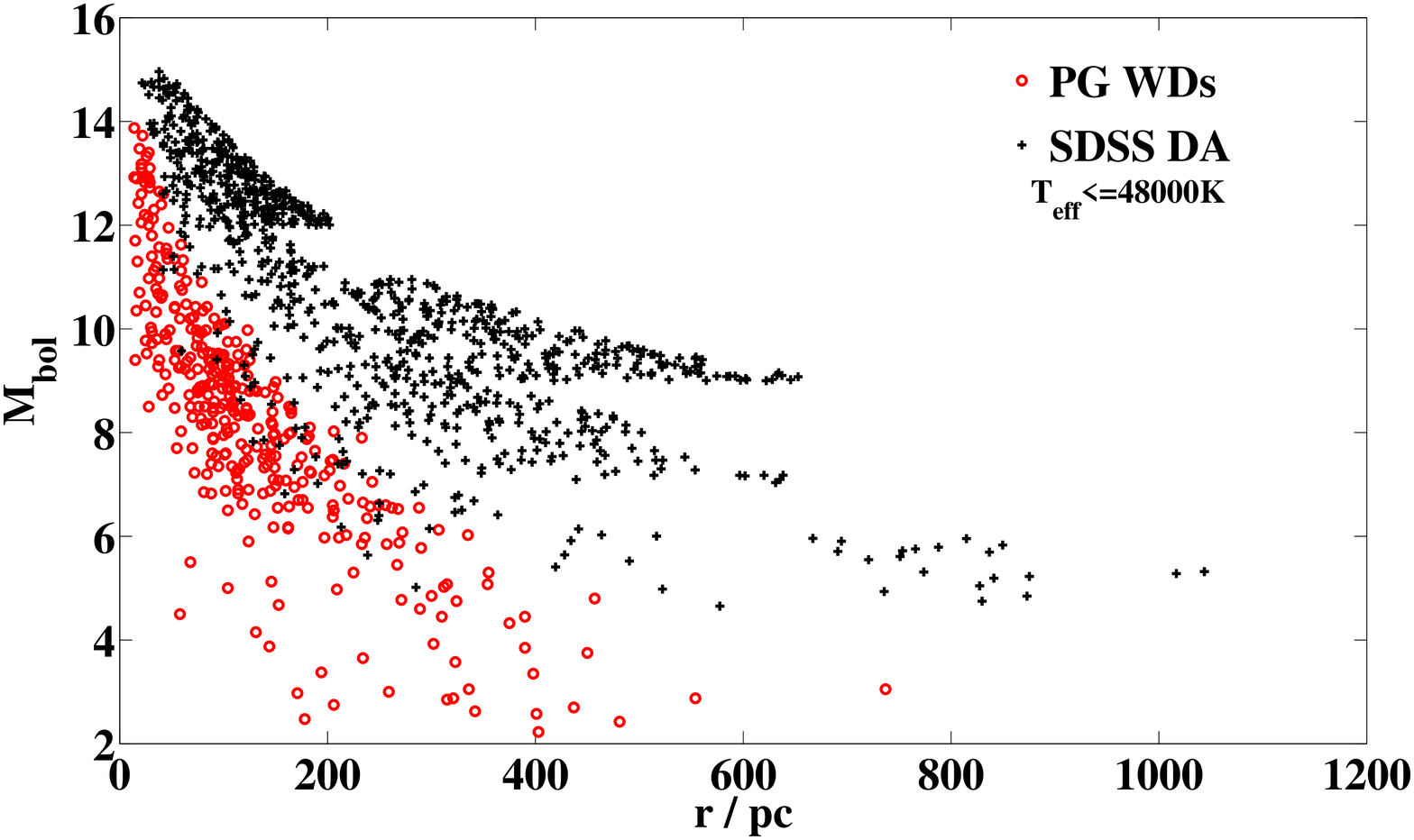}
    \caption{ Selection effect comparison. Dots denote PG DA WDs from Liebert
et al. (2005) and crosses the SDSS DA WDs after bin-correction. It is clear
that in the case of $M_{bol}<8$ mag  the PG sample is obviously more dense than
the SDSS sample, especially when considering the very large value of
$V_{max}-V_{min}$ of SDSS at the brighter magnitude end.}
\label{Fig.8}
\end{center}
\end{figure}

\subsection{LF of non-magnetic DA WDs and comparisons with previous results}

Fig. 6 shows the LF of SDSS non-magnetic DA WDs (solid line)
with $T_{eff}<48000$K and $M_V<15.25$. We also show other LFs
obtained previously: Liebert et al. (2005, dashed line, called the
PG sample and we select WDs with $T_{eff}<48000$K in their
sample for comparison); Green (1980, dotted line); Fleming et al.
(1986, dot-dashed line). Briefly, the SDSS LF is in general
agreement with Liebert et al (2005) and Fleming et al (1986),
especially at the fainter end ($M_V>12$). An obvious advantage of the
SDSS LF is that it extrapolates the fainter end  of WD LF from
13.25 to 15.25 mag. Because  the SDSS has a lower magnitude
limit, WDs with larger $M_V$ are usually nearer to 
us. Their selection effects may
not be very strong and the result should be less biased. At the
brighter end, however, the SDSS LF is lower by about half an order
of magnitude than the PG LF. Some possible reasons that account for 
such a difference:
(1) The PG Survey is an all-sky survey, whereas the SDSS DR1 just
covers 1360 $deg^2$ of the whole sky.
(2) The SDSS sample may contain fewer WDs at the brighter end
where $M_{bol}$ $<$7.5. Because SDSS has a low magnitude limit,
very bright WDs (with smaller $M_V$) must be 
very distant from us, as
shown in Fig. 8. These stars are extremely contaminated by the
selection effects and even after the bin-correction the result still may
be inaccurate. (3) Fleming et al. (1986) and Liebert et
al. (2005) both pointed out that there may be problems of missing
binaries or double degenerates in the PG WD sample. Zuckerman \&
Becklin (1992) and Marsh, Dhillon \& Duck (1995) have shown that
many low mass DA candidates (usually hot and with low absolute
magnitudes) are binaries, with the companion being either a low
mass main sequence star or another WD. Kleinman et al. (2004) also
mentioned this problem. Bergeron, Leggett \& Ruiz (2001) made a
detailed analysis of this unresolved problem. Liebert et al.
(2005) even pointed out that  double degenerates are likely in
the majority of cases. So we expect that the missing binaries in
the SDSS sample may account for a considerable number of missing stars.

\begin{figure}
    \begin{center}
    \includegraphics[width=9.2cm, height=7.5cm]{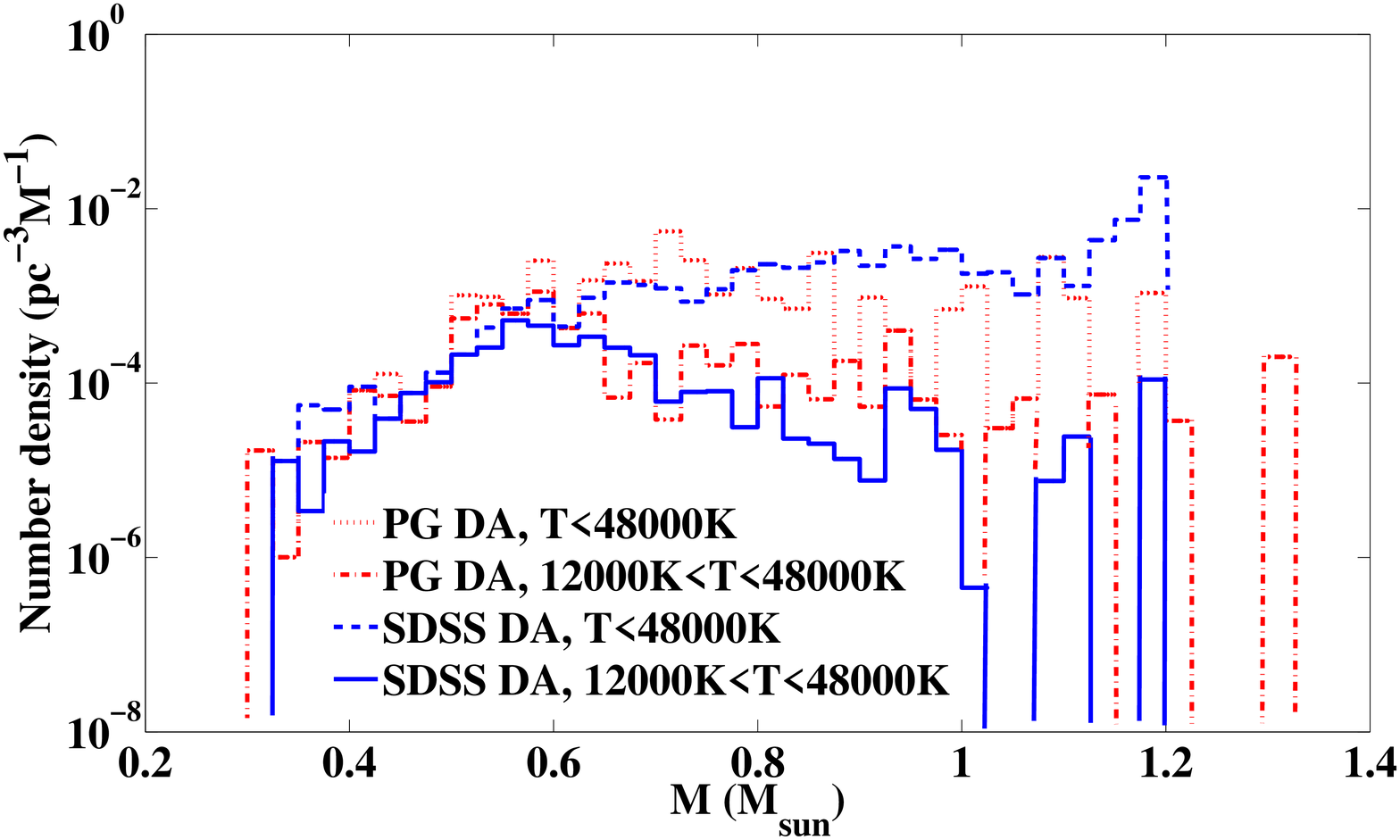}
    \caption{ Mass function of SDSS DA WDs in different $T_{eff}$ intervals and
the comparison with the PG sample (Liebert et al. 2005). The
dot-dashed and dotted lines denote PG DA samples with $12000K<T_{eff}<48000K$
and $T_{eff}<48000K$ from Liebert et al. (2005). The solid and
dashed lines denote SDSS DA samples with $12000K<T_{eff}<48000K$ and
$T_{eff}<48000K$, respectively. Note that $y$ axis of this figure is a
logarithmic scale.}
\label{Fig.9}
\end{center}
\end{figure}

\subsection{Comparison with theoretical works}

The SDSS DA WD LF is consistent with the theoretical predictions of 
Mestel (1952) and Lamb \& van Horn (1975). The
Mestel law is:
$
Log \phi  \propto  - (5/7) Log(L/L_\odot) .
$
 The evolutionary tracks obtained by Lamb \& van Horn (1975) also agree 
with the Mestel law in the
$M_{bol}$ range between 6.0 and 13.5. They explained that the
deviation below this range is due to the neutrino energy losses and
above this range due to the Debye cooling. Compared with
these, the SDSS DA WD LF is approximately a
straight line in the range of $M_{bol}$ between 6.0 and 13.5, with
a linear fitting slope of 0.32, which is almost identical to the
slope 2/7 of the Mestel law (Fig. 7). The SDSS DA WD LF also shows a
trend of deviation when $M_{bol}$ is smaller than 6.0, which is
identical with the model of Lamb \& van Horn (1975). For the fainter end where
$M_{bol}>13.5$, the LF data does not cover a sufficiently broad
$M_{bol}$ range to test the model of  Lamb \& van Horn (1975).

\section{Mass function and space density of SDSS DA WDs}

\begin{figure}
    \begin{center}
    \includegraphics[width=9.2cm, height=7.5cm]{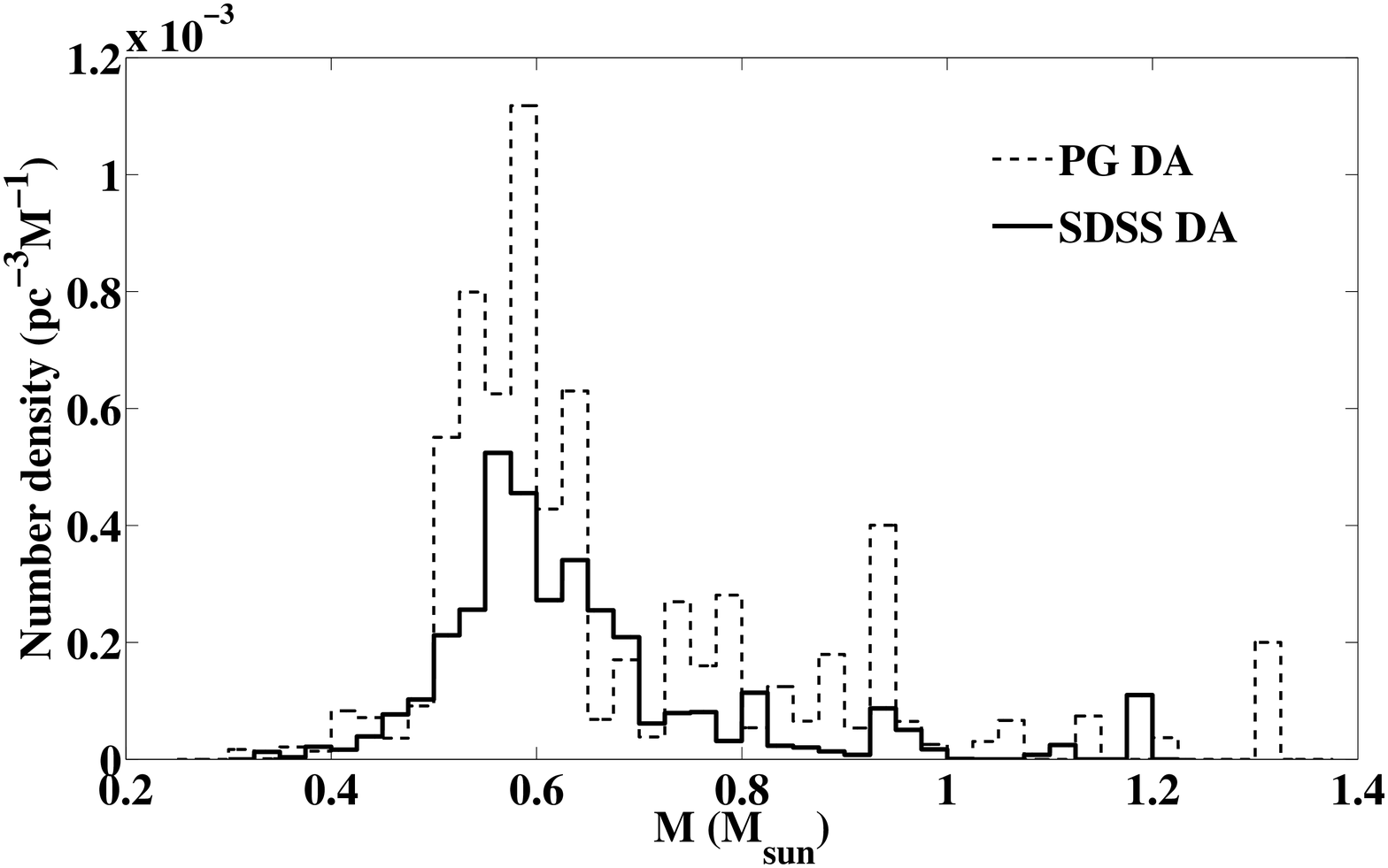}
    \caption{ Mass function of SDSS DA WDs (solid line) with $12000K<T_{eff}<48000K$ and the
 comparison with the 270 DA WDs in PG sample (dashed line) (Liebert et al. 2005).}
\label{Fig.10}
\end{center}
\end{figure}

Fig. 9 shows the mass function (MF), i.e. the
$1/(V_{max}-V_{min})$ weighted mass distribution of the SDSS DA
WDs. Kleinman et al. (2004) pointed out that the $\log g$ value
determination of cooler WDs with $T_{eff}<12000K$ has a systematic
offset to higher $\log\ g$ and a possible interpretation is that a
moderate amount of helium has been convectively mixed into the
atmosphere (see section 6 of Kleinman et al. 2004, also Bergeron et
al. 1990; BSL; Liebert et al. 2005). Thus, the parameters and
mass functions of WDs with $T_{eff}<12000K$ may be inaccurate.
However we include these cool WDs in our MF in Fig. 9 for
reference.  From both the PG and SDSS MFs in Fig. 9  we can see a
qualitative property of MF that in the massive part, the cool WDs'
space density is much larger than that of the hotter WDs. A
possible explanation is that the hot massive WDs usually evolve much
faster than cool WDs, which leads to its faintness (larger
$M_{bol}$) and  thus difficulty for observations. So the WDs we
observe are usually quite near to us, and consequently have a
small $r_{max}$ and larger $1/(V_{max}-V_{min})$ (see Fig. 5 and
Fig. 8), which will result in a higher space density. A
rough estimate leads to an important implication that cool
massive WDs may contribute a larger part to the galactic  matter
than previous estimates. However, the confirmation of this
requires further investigation with more accurate $log\ g$ measurements.

\subsection{Non-magnetic DA WD Mass Function}

Fig. 10 shows the usually discussed mass function (MF) of WDs with $T_{eff}$
between 12000K and 48000K. It is more accurate  because
we have more reliable estimates of the masses of these WDs 
(see discussions in section 2). In
many bins, the SDSS DA density is lower than the PG DA density,
and the reason is similar to those explained in subsection
4.1. However, their relative distributions are similar. The SDSS MF is also
similar to other previous studies, e.g. Wiedemann \& Koester
(1984), McMahan (1989), BSL, Marsh et al. (1997a), Vennes et
al. (1997), Finley et al. (1997) and Napiwotzki et al. (1999),
etc. The majority of WDs clump between 0.5 and 0.7 $M_{\odot}$.
with some small clusters from 0.7 to 1.0 $M_{\odot}$. Another
peak is perhaps seen at 1.2 $M_{\odot}$. Nevertheless, since Kleinman et
al. (2004) have artificially assigned an upper $\log g$ limit of
9.0, we obtain no WDs with mass higher than about 1.2 $M_{\odot}$.
In other words, the 1.2 $M_{\odot}$ cluster probably includes some WDs more
massive than 1.2 $M_{\odot}$. For this reason, Madej et al. (2004)
concluded that this peak is not a real feature. Since our sample
has been corrected for selection effects and is more complete, we
conclude that there really is a cluster and a peak around 1.2
$M_{\odot}$, while the peak may be slightly larger than 1.2
$M_{\odot}$.

\subsection{Continuous Mass Function}

\begin{figure}
    \begin{center}
    \includegraphics[width=9.2cm, height=7.5cm]{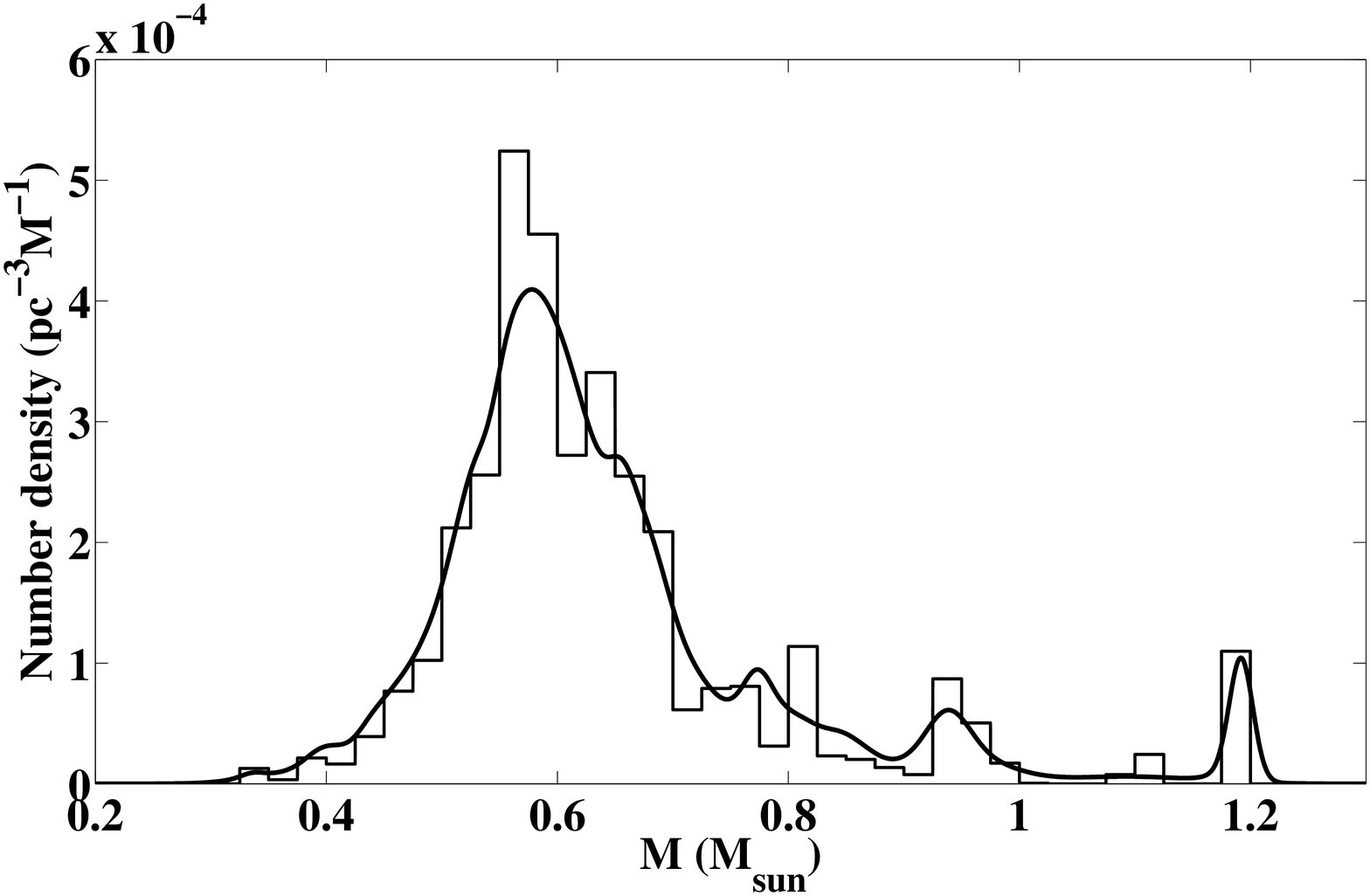}
    \caption{Comparison of the traditional discrete MF and our improved
continuous MF of 531 SDSS DA WDs with $12000K<T_{eff}<48000K$}
\label{Fig.11}
\end{center}
\end{figure}

\begin{figure}
    \begin{center}
    \includegraphics[width=9.2cm, height=7.5cm]{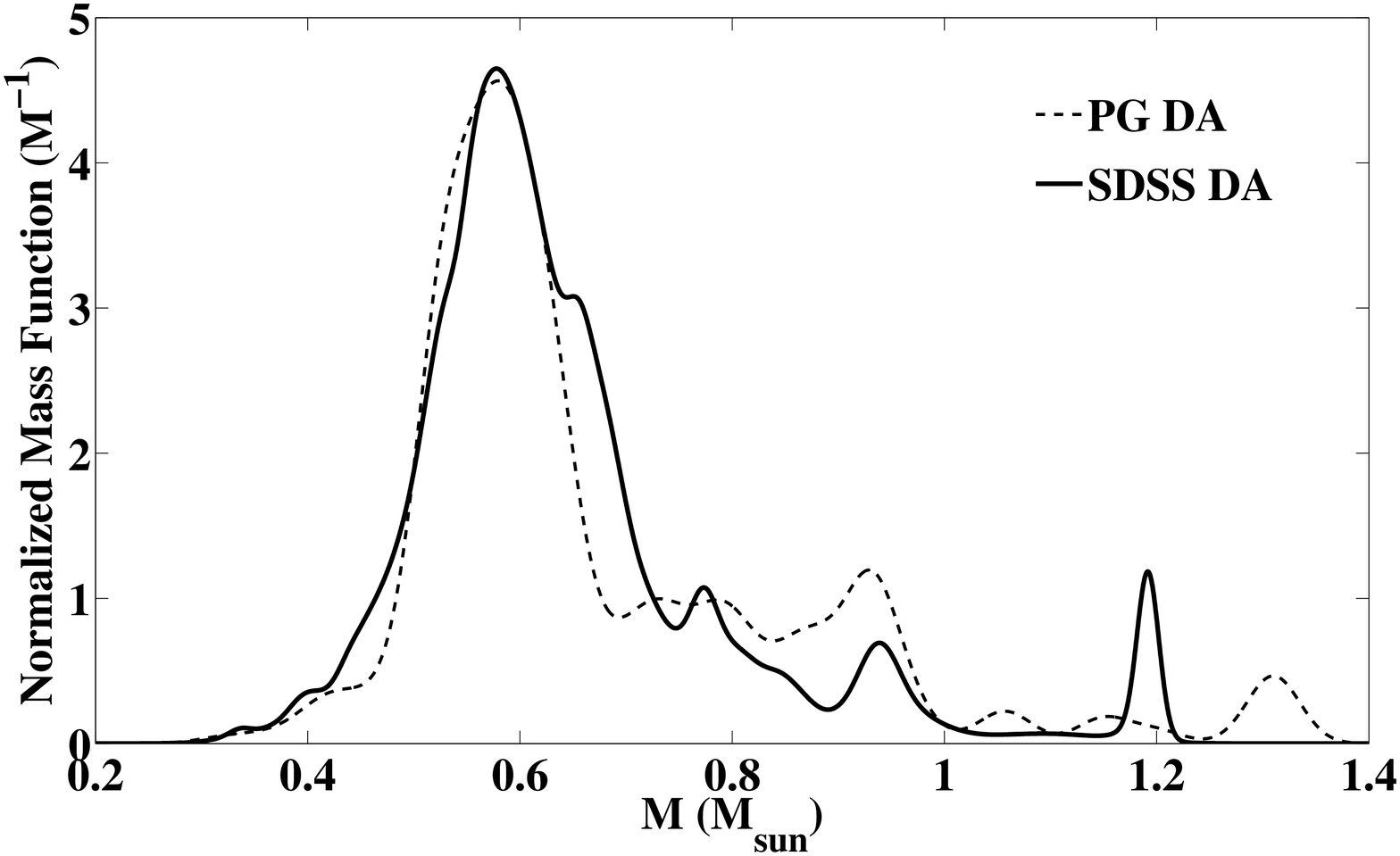}
    \caption{Normalized MF of 531 SDSS  DA WDs (solid line) compared to 270 PG
DA WDs (Liebert et al. 2005, dashed line), both with $12000K<T_{eff}<48000K$.}
\label{Fig.12}
\end{center}
\end{figure}

Vennes et al. (1997) described a method to derive a continuous MF by
calculating $dN_{<M}/dM$, where $N_{<M}$ denotes the number of WDs with 
mass less than a value M. At each mass point of a WD in the
sample, this $dN_{<M}/dM$ will result in a Dirac $\delta$
function. For this reason, they smoothed the function by
assuming a Gaussian distribution with a uniform FWHM of 0.1
$M_{\odot}$. Here we try to improve their method. For a specific WD with
its observation-derived mass $M_i$ and error $\sigma M_i$, the
probability density function of this WD is a Gaussian distribution
function. The probability that this WD's mass equals M is equal
to $\Psi _i (M)dM = \frac{1}{{\sqrt {2\pi } \sigma _{Mi} }}\exp
\{  - \frac{{(M - M_i )^2 }}{{2\sigma _{Mi}^2 }}\} dM $ and $\int
{\Psi _i (M)dM}  = 1$. So the mass function here can be defined as:
$$
\rho (M) = \sum\limits_i {\frac{1}{{V_{\max i}  - V_{\min i} }}
 \Psi _i (M)}~~~~~~~~~~~~~~~~~~~~~~~~~~~ $$
\begin{equation} 
~~~=\sum\limits_i {\frac{1}{{V_{\max i} - V_{\min i} }}
\frac{1}{{\sqrt {2\pi } \sigma _{Mi} }}\exp \{ - \frac{{(M - M_i
)^2 }}{{2\sigma _{Mi}^2 }}\} }.
\end{equation}
This is the detected or observed space density. The total
space density is $\rho _{tot}  = \sum\limits_i {\frac{1}{{V_{\max
i}  - V_{\min i} }}}$ and the normalized mass function will be:
$\rho _{norm} (M) = \rho (M)/\rho _{tot}$

Usually, the error of a SDSS WD mass is small enough to retain its
distribution properties and also large enough not to produce a
Dirac $\delta$ function. Fig. 11 compares the traditional discrete
MF and our continuous MF of SDSS DA WDs with $T_{eff}$ between
12000K and 48000K. They are in good agreement and thus demonstrate
the reliability of our method. The continuous MF has many
advantages over the discrete one. From it, we determine that the
main peak of the SDSS DA mass distribution is at M=0.58
$M_{\odot}$ and two other obvious peaks at M=0.94 $M_{\odot}$ and
M=1.19$M_{\odot}$. The 0.58$M_{\odot}$ peak is in perfect
agreement with previous studies (see Table 1 of Madej et al.
2004). The
0.58$M_{\odot}$ peak, which is derived from $1/(V_{max}-V_{min})$
weighted MF of a complete sample after selection effect
corrections, is close to the 0.562$M_{\odot}$ peak derived by
Madej et al. (2004), found by simply counting the number of
WDs in an incomplete sample. This implies that the main peak
of the WD mass distribution around 0.57 $M_{\odot}$ is very
insensitive to sample completeness, which puts in context the
agreement of our results with previous studies.

\subsection{Total space density and normalized mass function}

\begin{figure}
    \begin{center}
    \includegraphics[width=9.2cm, height=7.5cm]{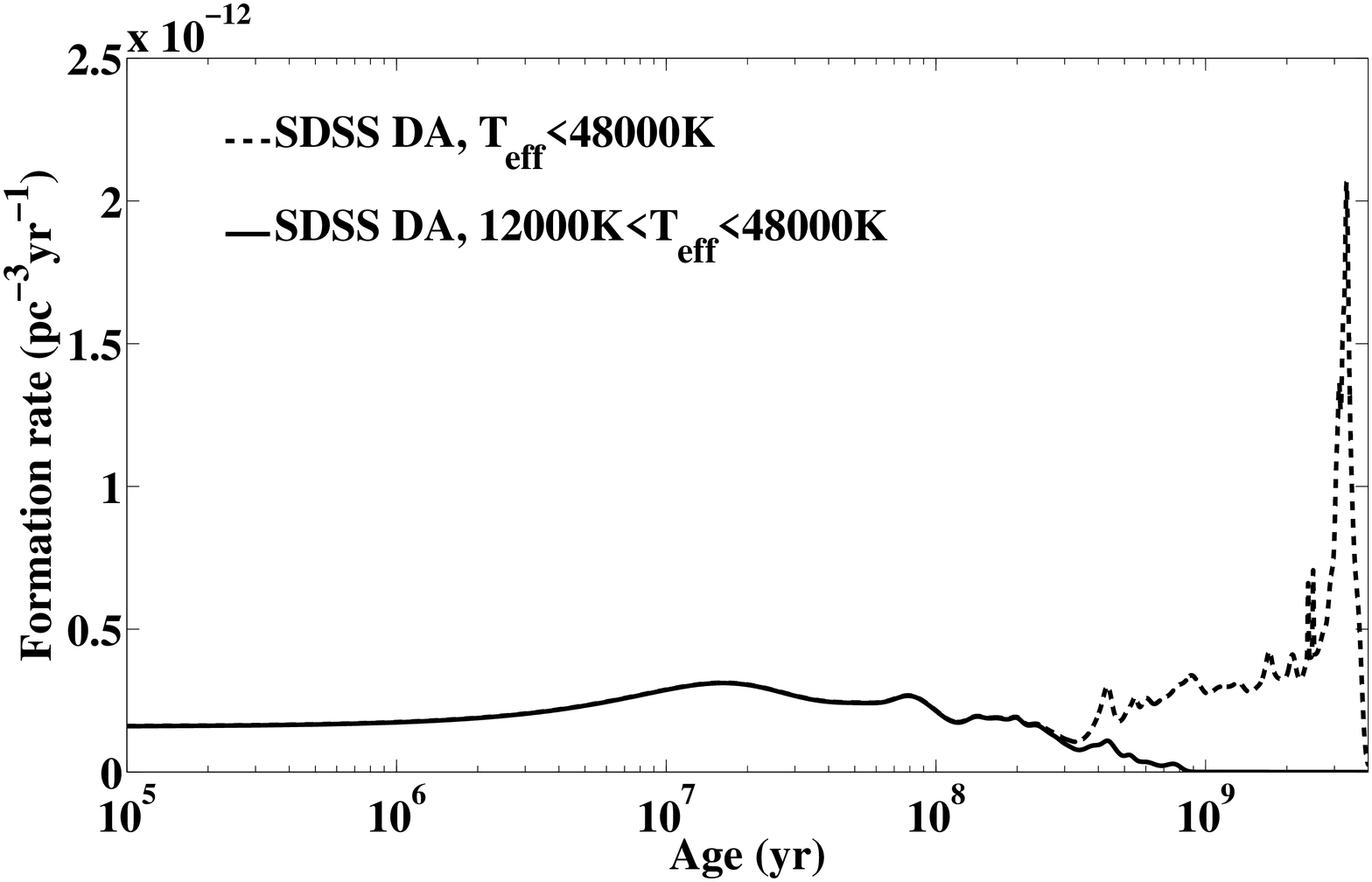}
    \caption{Detailed formation rates of SDSS WDs in linear and log Age scale.
 The dashed lines denote SDSS DA samples with $T_{eff}<48000K$. The solid lines denote SDSS DA samples with
$12000K<T_{eff}<48000K$ .} \label{Fig.13}
\end{center}
\end{figure}

The total space density of SDSS DA WDs is $8.81\times 10^{-5}
pc^{-3}$ for WDs with $T_{eff}$ between 12000K and 48000K and
$1.94\times10^{-3} pc^{-3}$ for WDs with $T_{eff}<48000K$. If we
include DB/DO WDs, the result will be $1.10\times10^{-4} pc^{-3}$
and $2.51\times10^{-3}$ pc-3, respectively. The normalized MF is
shown in Fig. 12, assuming that the average error for the PG DA WD masses is
0.025 $M_{\odot}$, equal to the bin-width of the discrete MF
to retain the distribution information. The two normalized MFs
agree well and show similar properties of distribution when $M< 1.0
M_{\odot}$, e.g. the main peak around 0.57 $M_{\odot}$  and its width
(FWHM), despite the SDSS DR1 just covering a small area of the
whole sky. As we discussed in section 6.2, if the artificial $\log
g =9.0$ limit is relaxed, the SDSS 1.2 $M_{\odot}$ high
and thin peak would be lower and wider and move right-ward,
more like  the PG 1.3 $M_{\odot}$ peak.
Our conclusion is that there is a small peak around or
above 1.2 $M_{\odot}$.

\section{The formation rate of SDSS WDs}

\begin{figure}
    \begin{center}
    \includegraphics[width=9.2cm, height=7.5cm]{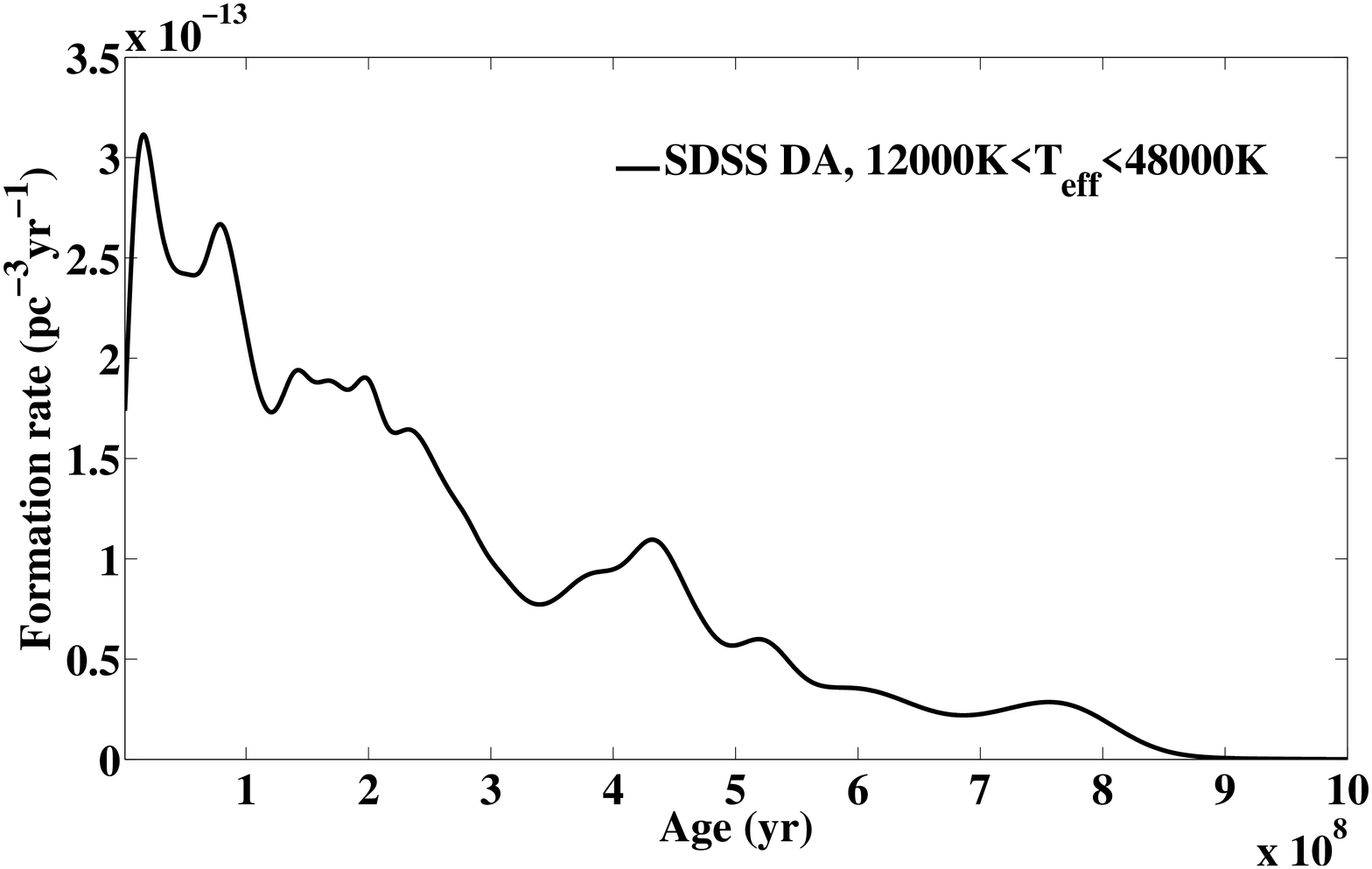}
    \caption{Detailed formation rates of SDSS WDs with $12000K<T_{eff}<48000K$.}
\label{Fig.14}
\end{center}
\end{figure}

\begin{figure}
    \begin{center}
    \includegraphics[width=9.2cm, height=6.5cm]{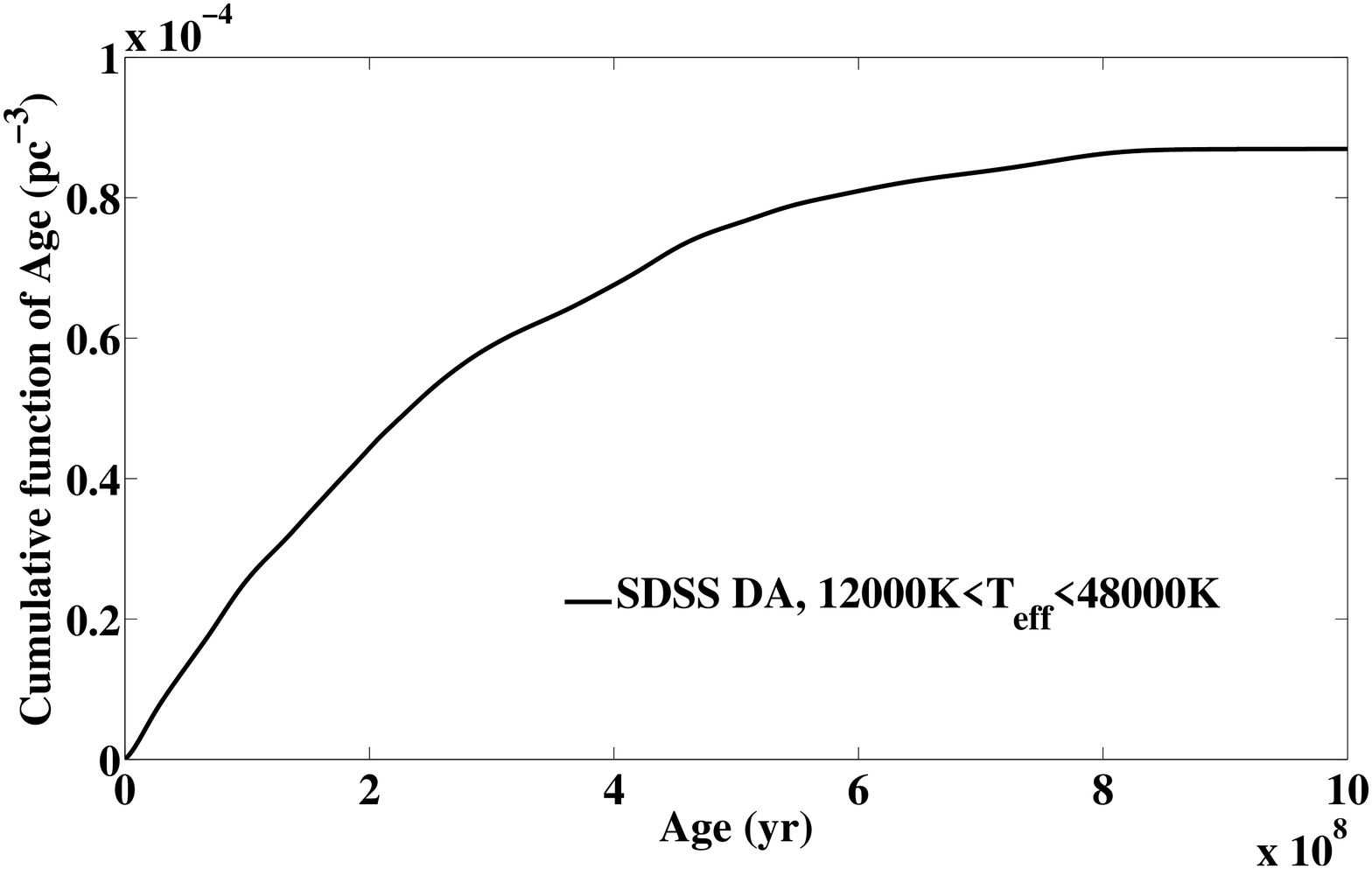}
    \includegraphics[width=9.2cm, height=6.5cm]{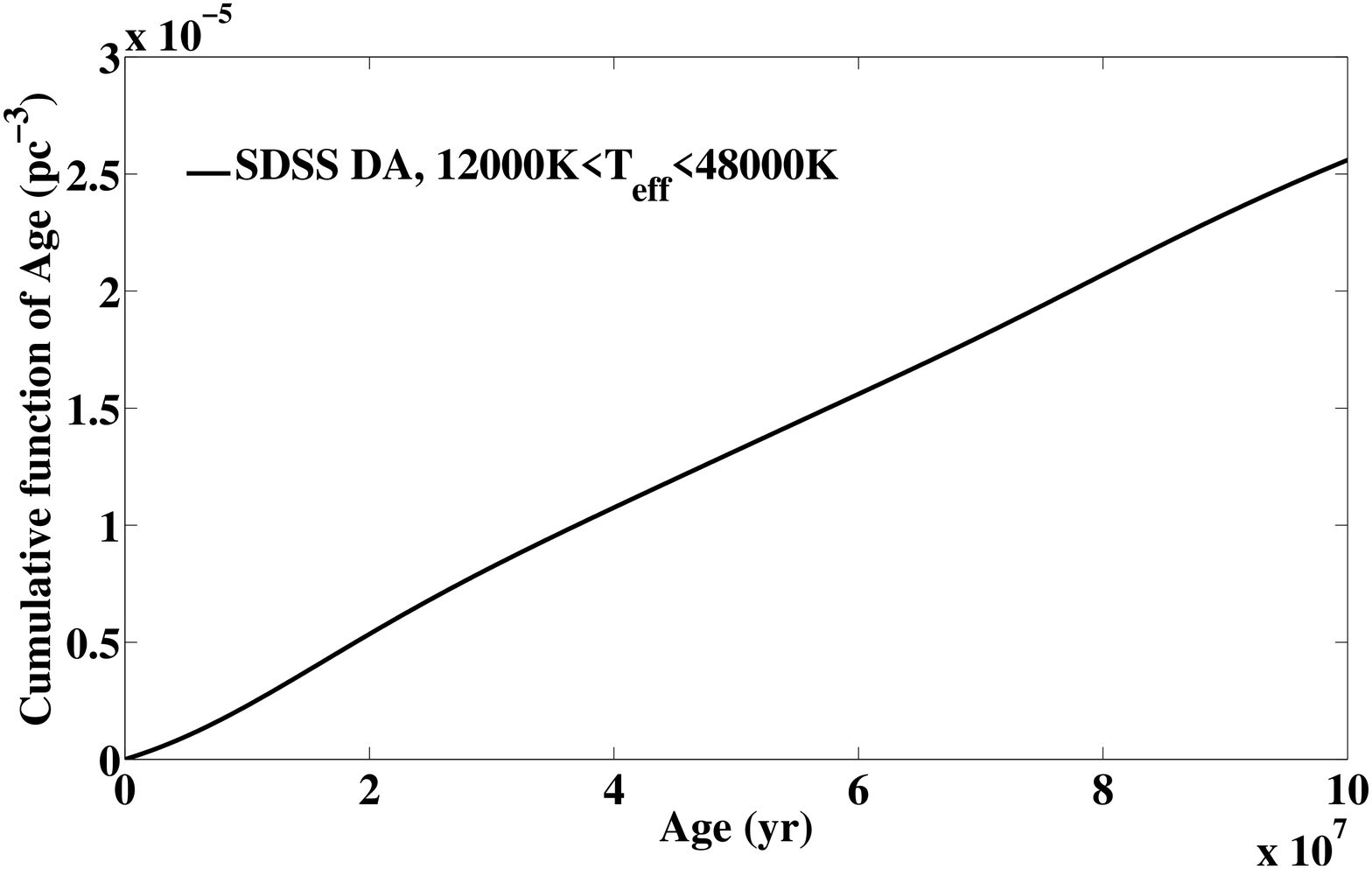}
    \caption{Cumulative Age Functions of SDSS WDs with
$12000K<T_{eff}<48000K$ in the last 1 Gyrs (upper panel) and 0.1 Gyrs 
(lower panel). It clearly shows a declining
slope with age, which implies a decline of the formation rate.
Yet in the last 0.1 Gyrs, the
curve is nearly a straight line.} \label{Fig.15}
\end{center}
\end{figure}

\begin{figure}
    \begin{center}
    \includegraphics[width=9.2cm, height=7.5cm]{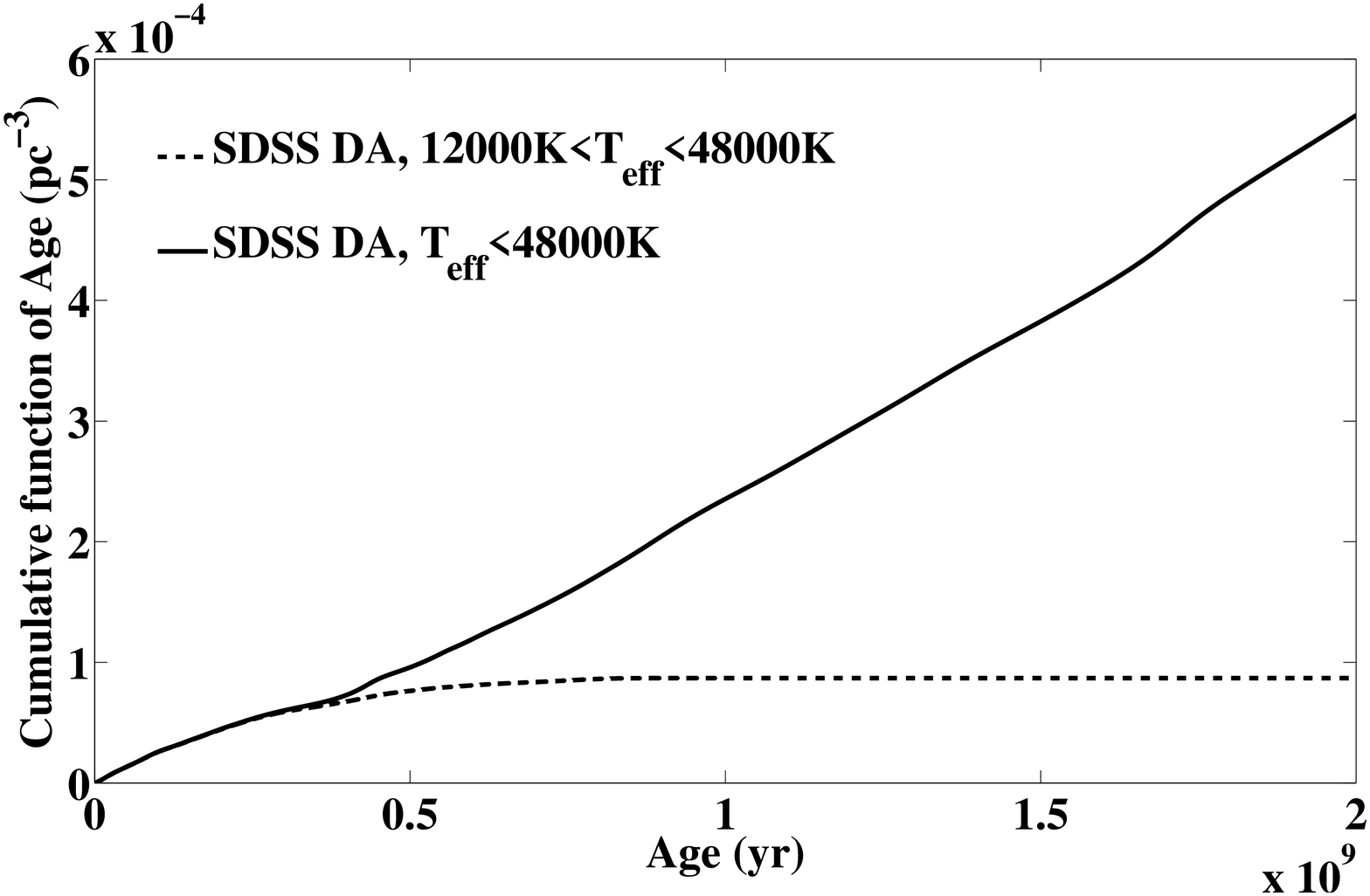}
    \caption{Cumulative Age Functions of SDSS WDs with $T_{eff}<48000K$. The
dashed lines denote SDSS DA samples with $T_{eff}<48000K$. The
solid lines denote SDSS DA samples with $12000K<T_{eff}<48000K$.
Dashed curves in the upper figure show variable slope as age goes
old, which implies the variation of the formation rate. Yet in the
last 2 Gyrs, the curves are
nearly two straight lines.} \label{Fig.16}
\end{center}
\end{figure}

When we substitute WD age for the mass $M$ in the continuous MF,
we will obtain the age Function, as shown in Fig. 13.
 In fact, this age function is just the WD
formation rate as a function of Age (see Fig. 14). Although the
subtle details will be contaminated by fluctuation, the general
trend is much more reliable. Considering the complete sample
with $T_{eff}<48000K$, including 893 WDs, we find a relatively
constant formation rate of about $0.3\sim0.4\times10^{-12} pc^{-3}
yr^{-1}$ during the last 2 Gyrs. There is a very high formation
rate peak around $3\sim3.5$ Gyr (see dashed line in Fig. 13). The 
real peak may not be as high as is
shown because of the artificial upper limit of
$\log g = 9.0$ by Kleinman et al. (2004). The massive WDs,
according to current SDSS data and models, are usually very old.
If we consider only the 531 WDs with $12000K<T_{eff}<48000K$,
another problem is seen: the formation rate declines rapidly
when the WD's age exceeds 0.1 Gyr (see Fig. 14). Liebert et al. (2005) noticed
this fact as well (see section 5.1 and Fig. 16 of their paper).
An interpretation can be found: when we just consider the
hotter WDs, the sample will be incomplete due to the elimination of the
cooler ones which were much hotter many years ago when they were
born. So the formation rate of the hotter sample will decline
rapidly accompanying the WDs' cooling along time. The same thing
happens in the EUVE Survey sample (see Fig. 9 in Vennes et al.
(1997)). In their sample the formation rate declines even faster
(at 10 Myr) than  in our work and Liebert et al. (2005)
because their sample is hotter than 20000K. The hotter the sample
is, the more incomplete it is and the more rapidly the formation
rate declines. 
We integrate the Age (birth rate) Function to get the Cumulative Age 
Function as shown in Fig. 15 and Fig. 16, which are similar to Fig. 9 
in Vennes et al. (1997). In Fig. 15, the hotter sample shows a nearly 
excellent straight line below 0.1 Gyr. By assuming a constant formation 
rate in the last 2 Gyrs to eliminate the influence of the fluctuation in 
the continuous function, we can  make a linear fit to obtain an 
average formation rate equal to the fitting line's slope. The 
complete sample also exhibits a
straight line below 2 Gyr, as is shown in Fig. 16. The result is
$2.579\times10^{-13} pc^{-3} yr^{-1}$ and $2.794\times10^{-13}
pc^{-3} yr^{-1}$, for WDs with $12000K<T_{eff}<48000K$ and
$T_{eff}<48000K$, respectively. If the formation rate is corrected
for nondegenerate companions and for those WDs that are likely to
be in binaries, it will be increased by a significant factor.
Compared with the recent calculated PNe formation rate of about
$(2.1\sim3) \times10^{-12} pc^{-3} yr^{-1}$ (Pottasch 1996;
Phillips 2002), there is a significant disagreement (see the detailed
discussion in section 5.7 of Liebert et al. 2005). Previous
calculated WD formation rates have also been listed in Table 3 for
comparison with our result.

\section{Three-dimension distribution function and the H-R diagram}

Bergeron et al. (2001) have emphasized the importance of combining
the MF and LS into one distribution function for the comparison of
cooling time. The LF and other parameters of a sample of WDs 
vary in different mass groups. For this reason, Liebert et al.
(2005) divided the whole PG sample into 3 groups with mass around
0.6 $M_{\odot}$, $M<0.46 M_{\odot}$ and $M>0.8M_{\odot}$,
respectively. Their discussions of the other parameters such as
the formation rate are all based on this 3-group division.
However, this method also has some problems. (1) Even within each
group, the distribution and parameters are not uniform. (2)
This 3-group division method is not universal, e.g. in the SDSS
sample, as is shown in Fig. 10 and Fig. 12, this division is not
appropriate because the 0.578 $M_{\odot}$ peak is too strong, 
overwhelming the other groups and the 1.19 $M_{\odot}$ peak is 
unreliable. Using the continuous distribution function we proposed
above, a more universal solution can be found by assuming that
every WD's contribution to the space density can be described as
a 2-Dimensional Gaussian distribution weighted by
$1/(V_{max}-V_{min})$. We define the Mass-Luminosity Function
(MLF) as:
$$
 \rho (M,M_{bol} ) =  \\
 \sum\limits_i {(\frac{1}{{V_{\max ,i}  - V_{\min ,i} }} \cdot \frac{1}{{2\pi \sigma _{M_i}  \cdot \sigma _{M_{bol,i}} }} \cdot }  $$
\begin{equation}
~~~~~~~~~~~~~~~ \exp \{  - \frac{{(M - M_i )^2 }}{{2\sigma _{M,i}^2 }} - \frac{{(M - M_{bol,i} )^2 }}{{2\sigma _{M_{bol,i}}^2 }}\} ) \\
\end{equation}
One can substitute the age, $T_{eff}$ or other parameters for $M$ or
$M_{bol}$ in Eq. (4) to obtain other 3-D distribution functions.

\begin{figure}
    \begin{center}
    \includegraphics[width=9.2cm, height=7.5cm]{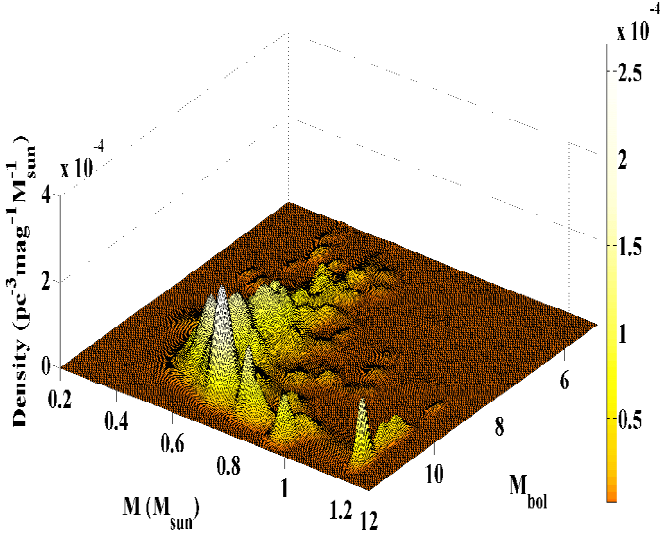}
    \caption{3-Dimension Mass-Luminosity Function for SDSS DA WDs with $12000K<T_{eff}<48000K$.}
\label{Fig.17}
\end{center}
\end{figure}

\begin{figure}
    \begin{center}
    \includegraphics[width=9.2cm, height=7.5cm]{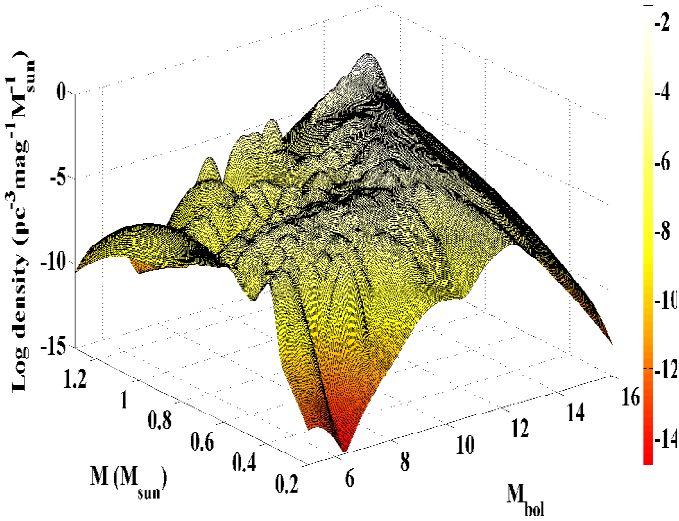}
    \caption{3-Dimension Mass-Luminosity Function for SDSS DA WDs with $T_{eff}<48000K$. Note
    that the z axis is in log scale. The red part which has the highest value shows the trend
    that the mass peak moves toward the high mass end as $M_{bol}$ becomes higher.}
\label{Fig.18}
\end{center}
\end{figure}

\begin{figure}
    \begin{center}
    \includegraphics[width=9.2cm, height=7.5cm]{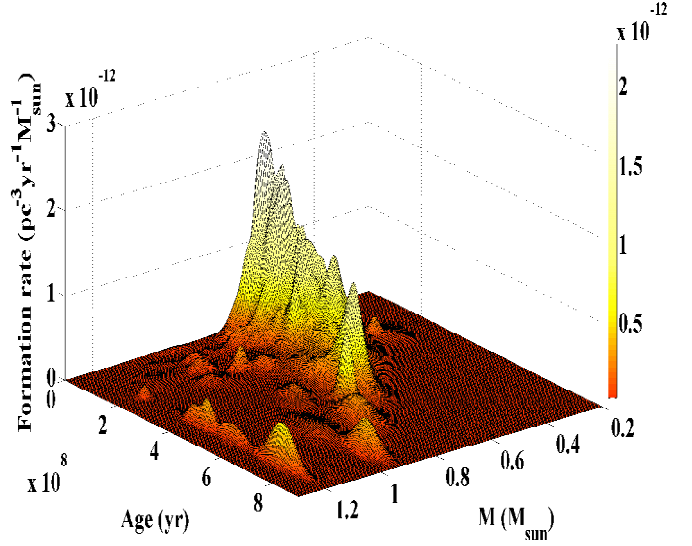}
    \caption{Mass-age distribution function for SDSS DA WDs with 
$12000K<T_{eff}<48000K$ (upper) and $T_{eff}<48000K$ (lower). The mass peak 
has a explicit trace toward the high mass end when the age increases.}
\label{Fig.19}
\end{center}
\end{figure}

\begin{figure}
    \begin{center}
    \includegraphics[width=9.2cm, height=7.5cm]{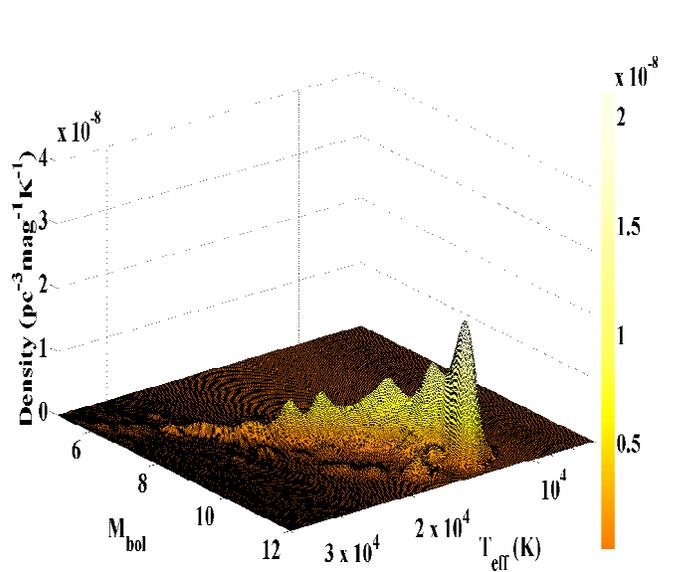}
    \caption{H-R diagram for SDSS DA WDs with $12000K<T_{eff}<48000K$.  $T_{eff}$ axis is in $\log$ scale and the left end has a larger value.}
\label{Fig.20}
\end{center}
\end{figure}

Fig. 17 shows the mass-luminosity function for SDSS DA WDs with
$12000K<T_{eff}<48000K$. The main cluster in Fig. 12 with many
subtle peaks is now decomposed into a 'mountain' along
$M_{bol}$ with several independent peaks. In this figure, the
maximum mass distribution peak is at about 0.65 $M_{\odot}$, which
is just a quasi-peak in Fig. 12, and the overwhelming peak at 0.58
$M_{\odot}$ is the integration along $M_{bol}$ of the
'mountain ridge' at about 0.55 $M_{\odot}$. We can also see a
trend that the mass cluster moves to the massive end as the
$M_{bol}$ goes to the fainter end. Thus, we can infer that if the
$M_{bol}$ becomes fainter than 12 mag, the mass cluster will continue
to move to the massive end, and in Fig. 18 we confirm this inference.
This trend is partly interpreted by Fig. 19. If we assume that most
of the progenitors of these WDs formed almost simultaneously at
early stage of our galaxy, as massive stars usually evolve
faster than lighter ones, they will soon die out to produce
massive WDs and have a long cooling time to reach a present high
$M_{bol}$. Meanwhile the less massive stars have much longer lives and
die slowly and much later to give birth to recently born WDs,
which have had little time to cool and only retain a low $M_{bol}$. This
implies that the mass - age distribution of a complete WD sample
contains important information on the early main sequence stars and
our galaxy. Fig. 20 gives the 3-D H-R diagram which shows an
obvious evolutionary track of WDs.

\section{Discussion and summary}

We have performed a study of a DA WD
sample from SDSS DR1. To ensure that our adopted sample is
accurate and complete, we have carried out many tests. We
performed a mass determination comparison to test the accuracy of the
model of Panei et al. (2000). By comparing the model-derived mass with
that
 obtained from other methods, independent of the theoretical M-R
relation, we find that the model of Panei et al. (2000) is
reliable enough to be applied to the mass estimation of SDSS WDs,
especially for the hotter WDs.

We tested the completeness of the SDSS WD sample
and corrected for selection effects. We found 
that this sample is far from complete, mainly due to
the magnitude-limiting selection effect. Thus we lower the upper
magnitude limit by at least 1.5 mag to make the sample almost
complete. We also proposed a more detailed bin-correction method
to improve the accuracy. The remaining 531 DA WDs with
$12000K<T_{eff}<48000K$ still form the  largest homogeneous
and complete DA WD sample to date.

We calculated the SDSS WD Luminosity Function based on the
methods of Green (1980) and Fleming et al. (1986), except for some
minor differences. The SDSS LF is generally in agreement with most
previous studies. In addition, the SDSS LF itself shows excellent
 agreement with theoretical work, e.g.
the Mestel law. We also proposed some possible interpretations to
account for the disagreements  between the SDSS and PG WD
samples. We then introduced an improved continuous mass function
and a method to obtain the 2-dimensional distribution functions. We
derived a SDSS MF whose relative distribution and properties are
in good agreement with that of the PG sample, although the
absolute value is different. We thus obtained a 0.58 $M_{\odot}$
mass peak and found that in the 2-D Mass-Luminosity Function it is decomposed
into a 'mountain ridge' along $M_{bol}$ at about 0.55 $M_{\odot}$
and an actual peak at about 0.65 $M_{\odot}$ for fainter WDs. 
Evidence implies that there is a massive
WD peak or cluster above 1.2 $M_{\odot}$, which mainly consists of
 cool and faint WDs. The derived space
density is $8.81\times10^{-5} pc^{-3}$ and the formation rate is
about $2.579\times10^{-13} pc^{-3} yr^{-1}$ and
$2.794\times10^{-13} pc^{-3} yr^{-1}$, for SDSS DA WDs with
$12000K<T_{eff}<48000K$ and $T_{eff}<48000K$, respectively.

As predicted by Kleinman et al. (2004), we can expect an
additional 10,000 WDs or so by the time the SDSS is finished; 
  a much larger and more complete
sample of WDs will be available. Eisentein et al.
(2006) published a catalog of 9316 spectroscopically confirmed WDs from 
the SDSS DR4, which includes 8000 DA WDs. A further study of the statistical
properties of WDs with this
enlarged sample is under way and will be reported in a future work.
The theortical model of WDs adopted in our
study is still quite simple. Any future progress in the theoretical
models of WDs would be very helpful for a more accurate understanding of
the statistical properties of WDs.

\begin{acknowledgements}
We are greatly indebted to Professor P.~Bergeron for kindly
providing us with the newly calculated WD bolometric correction data.
We thank the referee, Professor Martin~A.~Barstow, for his careful 
annotations on the manuscript, which greatly improve our presentation.
This research is supported by the President Fund of Peking
University, the NFSC grants (No. 10473001 and No. 10525313),  the
RFDP grant (No. 20050001026) and the Key Grant Project of
Chinese Ministry of Education (No. 305001).
\end{acknowledgements}

\onecolumn
\newpage
\begin{table}[t]
\caption[ ]{Comparison of white dwarf mass determinations from
different methods} 
\tabcolsep 11pt
\begin{tabular*}{168mm}{lcccccc}
\hline
Name& $T_{eff }$ &     $log\ g$ &  $M_{spec}$&      $M_{other}$&    Notes   & Ref.\\
&(K)&&($M_{\sun}$)&($M_{\sun}$)&&\\
 \hline
Sirius B    &   24700  $\pm$  300 &   8.61   $\pm$  0.04    &   0.998  $\pm$ 0.024   &   1.003  $\pm$  0.016   &   1   &   1,3 \\
40 Eri B    &   16700   $\pm$ 300 &   7.77    $\pm$ 0.01    &   0.493   $\pm$ 0.006   &   0.501   $\pm$ 0.011   &   1   &   1,4 \\
Sirius B    &   25193  $\pm$  37 &   8.566   $\pm$  0.01    &   0.978  $\pm$ 0.005   & 1.02  $\pm$  0.02   &   2   &   14 \\
GD279   &   13500   $\pm$ 200 &   7.83    $\pm$ 0.03    &   0.514   $\pm$ 0.016   &   0.44    $\pm$ 0.02    &   2   &   1,12    \\
Feige 22    &   19100   $\pm$ 400 &   7.78    $\pm$ 0.04    &   0.505   $\pm$ 0.020   &   0.41    $\pm$ 0.03    &   2   &   1,12    \\
EG21    &   16200   $\pm$ 300 &   8.06    $\pm$ 0.05    &   0.649   $\pm$ 0.030   &   0.5 $\pm$ 0.02    &   2   &   1,12    \\
EG50    &   21000   $\pm$ 300 &   8.10    $\pm$ 0.05    &   0.682   $\pm$ 0.030   &   0.58    $\pm$ 0.05    &   2   &   1,12    \\
GD140   &   21700   $\pm$ 300 &   8.48   $\pm$  0.05    &   0.917  $\pm$  0.031   &   0.79  $\pm$   0.02    &   2   &   1,12    \\
G226-29 &   12000   $\pm$ 200 &   8.29   $\pm$  0.03    &   0.784   $\pm$ 0.019   &   0.75   $\pm$  0.03    &   2   &   1,12    \\
WD2007-303  &   15200   $\pm$ 700 &   7.86    $\pm$ 0.05    &   0.534 $\pm$   0.028   &   0.44   $\pm$  0.05    &   2   &   1,12    \\
Wolf1346    &   20000   $\pm$ 300 &   7.83    $\pm$ 0.05    &   0.532   $\pm$ 0.026   &   0.44    $\pm$ 0.01    &   2   &   1,12    \\
G93-48  &   18300   $\pm$ 300 &   8.02  $\pm$   0.05    &   0.630  $\pm$  0.029   &   0.75   $\pm$  0.06    &   2   &   1,12    \\
L711-10 &   19900  $\pm$  400 &   7.93   $\pm$  0.05     &  0.584  $\pm$  0.028   &   0.54   $\pm$  0.04    &   2   &   1,12    \\
CD-38 10980 &   24000  $\pm$  200 &   7.92  $\pm$   0.04    &   0.588 $\pm$   0.021   &   0.74   $\pm$  0.04    &   2   &   1,12    \\
Wolf 485A   &   14100  $\pm$  400 &   7.93  $\pm$   0.05    &   0.570  $\pm$  0.018   &   0.59   $\pm$  0.04    &   2   &   1,12    \\
G154-B5B    &   14000  $\pm$  400 &   7.71   $\pm$  0.05    &   0.457  $\pm$  0.015   &   0.53    $\pm$ 0.05    &   2   &   1,12    \\
G181-B5B    &   13600  $\pm$  500 &   7.79  $\pm$  0.05    &   0.495  $\pm$  0.016   &   0.46    $\pm$ 0.08    &   2   &   1,12    \\
G238-44 &   20200  $\pm$  400 &   7.90   $\pm$  0.05    &   0.568  $\pm$  0.028   &   0.42   $\pm$  0.01    &   2   &   1,12    \\
LB 1497 &   31660   $\pm$ 350 &   8.78    $\pm$ 0.049   &   1.097   $\pm$ 0.025   &   1.025  $\pm$  0.043   &   3,5 &   2,6,12  \\
HZ 4    &   14770  $\pm$  350 &   8.16    $\pm$ 0.049   &   0.707   $\pm$ 0.032   &   0.632   $\pm$ 0.042   &   3,5 &   2,7,12  \\
GH 7-112    &   15190  $\pm$  350 &   8.3 $\pm$ 0.049   &   0.795   $\pm$ 0.031   &   0.783   $\pm$ 0.039   &   3,5 &   2,7,12  \\
40 Eri B    &   16570  $\pm$  350 &   7.86   $\pm$  0.049   &   0.538   $\pm$ 0.026   &   0.52   $\pm$  0.4 &   3,5 &   2,8,12  \\
GH 7-191    &   19570   $\pm$ 350 &   8.09   $\pm$  0.049   &   0.674 $\pm$   0.030   &   0.669  $\pm$  0.036   &   3,5 &   2,7,12  \\
GH 7-233    &   24420  $\pm$  350 &   8.11   $\pm$  0.049   &   0.695  $\pm$  0.029   &   0.617  $\pm$  0.028   &   3,5 &   2,7,12  \\
HZ 7    &   21340  $\pm$  350 &   8.04   $\pm$  0.049   &   0.648  $\pm$ 0.029   &   0.665  $\pm$  0.077   &   3,5 &   2,7,12  \\
HZ 14   &   27390  $\pm$  350 &   8.07   $\pm$  0.049   &   0.678  $\pm$  0.028   &   0.51   $\pm$  0.086   &   3,5 &   2,7,12  \\
G191-B2B    &   64100  $\pm$  350 &   7.69  $\pm$  0.049   &   0.580  $\pm$  0.010   &   0.538  $\pm$  0.043   &   3,5 &   2,10,12 \\
G163-50 &   15070   $\pm$ 350 &   7.83   $\pm$  0.049   &   0.519   $\pm$ 0.026   &   0.465  $\pm$  0.046   &   3,5 &   2,5,12  \\
G148-7  &   15480  $\pm$  350 &   7.97   $\pm$  0.049   &   0.595  $\pm$  0.028   &   0.558  $\pm$  0.038   &   3,5 &   2,8,12  \\
Wolf 485A   &   14100  $\pm$  350 &   7.93   $\pm$  0.049   &   0.570   $\pm$ 0.028   &   0.529  $\pm$ 0.042   &   3,5 &   2,5,12  \\
L762-21 &   18580  $\pm$  350 &   8.32   $\pm$  0.049   &   0.813  $\pm$  0.032   &   0.808  $\pm$  0.099   &   3,5 &   2,5,12  \\
G154-B5B    &   13950   $\pm$    350 &   7.71     $\pm$   0.049   &   0.457    $\pm$   0.024   &   0.524     $\pm$  0.04    &   3,5 &   2,8,12  \\
G142-B2A    &   14040     $\pm$  350 &   7.84     $\pm$   0.049   &   0.521     $\pm$  0.026   &   0.561     $\pm$  0.037   &   3,5 &   2,8,12  \\
L587-77A    &   9330     $\pm$   350 &   7.87     $\pm$   0.049   &   0.524     $\pm$  0.028   &   0.657     $\pm$  0.035   &   3,4,5   &   2,5,12  \\
G86-B1B &   9140     $\pm$   350 &   8.3   $\pm$  0.049   &   0.785    $\pm$  0.032   &   0.454     $\pm$  0.118   &   3,4,5   &   2,11,12 \\
G111-71 &   7710      $\pm$  350 &   8.15     $\pm$   0.049   &   0.685    $\pm$   0.032   &   0.632     $\pm$  0.125   &   3,4,5   &   2,11,12 \\
G116-16 &   8750     $\pm$   350 &   8.29     $\pm$   0.049   &   0.778    $\pm$   0.032   &   1.009     $\pm$  0.06    &   3,4,5   &   2,11,12 \\
G121-22 &   10260     $\pm$  350 &   6.12      $\pm$  0.049   &   0.651    $\pm$   0.035   &   1.084     $\pm$  0.023   &   3,4,5   &   2,11,12 \\
G61-17  &   11000   $\pm$    350 &   8.04    $\pm$   0.049   &   0.626   $\pm$   0.030   &   0.552   $\pm$  0.038   &   3,4,5   &   2,11,12 \\
L619-50 &   10080  $\pm$   350 &   8.17   $\pm$  0.049   &   0.703 $\pm$   0.032   &   0.502  $\pm$   0.069   &   3,4,5   &   2,5,12  \\
LP696-4 &   10470   $\pm$  350 &   8.11   $\pm$  0.049   &   0.667  $\pm$  0.031   &   0.44   $\pm$   0.12    &   3,4,5   &   2,11,12 \\
LP25-436    &   8440    $\pm$   350 &   8.52   $\pm$   0.049   &   0.926  $\pm$   0.032   &   0.644   $\pm$   0.056   &   3,4,5   &   2,11,12 \\
G156-64 &   7160    $\pm$   350 &   8.43    $\pm$   0.049   &   0.866   $\pm$   0.032   &   0.548   $\pm$   0.052   &   3,4,5   &   2,11,12 \\
G216-B24B   &   9860    $\pm$    350 &   8.2   $\pm$  0.049   &   0.722    $\pm$   0.032   &   0.832     $\pm$  0.047   &   3,4,5   &   2,11,12 \\
L557-71 &   8780     $\pm$   350 &   8.29     $\pm$   0.049   &
0.778    $\pm$ 0.032   &   0.549    $\pm$   0.038   &   3,4,5   &
2,11,12\\
G19-20  &   13620  $\pm$  350 &   7.79   $\pm$  0.049   &   0.495   $\pm$ 0.025   &   0.489   $\pm$ 0.084   &   3,5 &   2,5,12\\

\hline
\smallskip
\end{tabular*}
\\
\emph{Note:} $M_{spec}$ refers to spectroscopic mass derived with the
evolutionary model of Panei(2000) from the spectral
parameters($T_{eff}$ and $log\ g$) and $M_{other}$ refers to the
mass derived from 1: orbital parameters, 2:
parallaxes and $log\ g$, and 3: gravitational redshift. Note 4 denotes low temperature
WDs ($T_{eff}$ $<$12000K) and 5 denotes that the individual errors of $T_{eff}$
and $log\ g$ of each WD is not available, so we assumed the errors of
$T_{eff}$ and $log\ g$ being the mean errors of the sample the WD
belongs to.

\emph{References:} (1) Provencal et al. (1998); (2) Bergeron et al. (1995a); 
(3) Gatewood \& Gatewood (1978); (4) Shipman et al. (1997); (5) Koester 
(1987) (6) Wegner, Reid \& McMahan (1991) (7) Wegner, Reid \& McMahan
(1989); (8) Wegner \& Reid (1987); (9) Koester \& Weidemann (1991); (10)
Reid \& Wegner (1988); (11) Wegner \& Reid (1991); (12) BSL and
Bragaglia et al. (1995); (13) McCook \& Sion (1999); (14) Barstow et al. (2005).
\end{table}

\newpage
\begin{table}[t]
\caption[ ]{Fundamental parameters of the DA white dwarfs from the Sloan Digital Sky Survey}
\tabcolsep 4pt
\begin{tabular*}{175mm}{lccccccccc}
\hline
Object name& SDSS No.  &            $T_{eff}$&        $log\ g$&        Mass &     Radius &      Age&    $M_{bol}$&  {$M_g$}&  Distance   \\
&&(K)&&($M_{\sun}$)&($\frac{R_{\sun}}{100}$)&(yr)&&&(pc)\\
\hline
002509	-	104048	&	1189	&	16390	$\pm$	480	&	7.74 	$\pm$	0.12 	&	0.48 	$\pm$	0.06 	&	1.54 	$\pm$	0.12	&	9.48E+07	&	9.27 	&	10.63 	&	555 	$\pm$	26  	\\
020747	+	121028	&	1210	&	16661	$\pm$	2104	&	8.37 	$\pm$	0.23 	&	0.84 	$\pm$	0.14 	&	0.99 	$\pm$	0.18	&	3.19E+08	&	10.16 	&	11.57 	&	371 	$\pm$	55  	\\
023907	+	002917	&	1198	&	16449	$\pm$	467	&	7.86 	$\pm$	0.11 	&	0.54 	$\pm$	0.06   &	1.43 	$\pm$	0.11 	&	1.14E+08	&	9.43 	&	10.80 	&	554 	$\pm$	24 	\\
024855	-	004139	&	1175	&	16180	$\pm$	390	&	7.87 	$\pm$	0.08 	&	0.54 	$\pm$	0.04 	&	1.42 	$\pm$	0.08 	&	1.24E+08	&	9.52 	&	10.84 	&	417 	$\pm$	14 	\\
025746	+	010106	&	1208	&	16628	$\pm$	214	&	8.30 	$\pm$	0.04 	&	0.80 	$\pm$	0.03 	&	1.05 	$\pm$	0.03 	&	2.78E+08	&	10.06 	&	11.46 	&	149 	$\pm$	3 	\\
			&		&		&		&				&				&		&		&		&				\\
030028	-	084126	&	1185	&	16364	$\pm$	505	&	7.90 	$\pm$	0.11 	&	0.56 	$\pm$	0.06 	&	1.39 	$\pm$	0.10 	&	1.25E+08	&	9.51 	&	10.87 	&	514 	$\pm$	23 	\\
031232	-	060907	&	1172	&	16164	$\pm$	373	&	7.86 	$\pm$	0.08 	&	0.54 	$\pm$	0.04 	&	1.43 	$\pm$	0.08	&	1.22E+08	&	9.51 	&	10.83 	&	438 	$\pm$	15  	\\
032126	-	061442	&	1187	&	16374	$\pm$	340	&	7.83 	$\pm$	0.07 	&	0.52 	$\pm$	0.04 	&	1.46 	$\pm$	0.07	&	1.11E+08	&	9.41 	&	10.76 	&	458 	$\pm$	14  	\\
032947	+	010050	&	1204	&	16590	$\pm$	726	&	8.59 	$\pm$	0.12 	&	0.98 	$\pm$	0.07 	&	0.83 	$\pm$	0.08 	&	4.52E+08	&	10.57 	&	11.97 	&	324 	$\pm$	20 	\\
032960	-	000733	&	1193	&	16406	$\pm$	326	&	7.76 	$\pm$	0.07 	&	0.49 	$\pm$	0.03 	&	1.52 	$\pm$	0.07 	&	9.83E+07	&	9.30 	&	10.66 	&	417 	$\pm$	12 	\\

\hline
\smallskip
\end{tabular*}
\\
\emph{Note:} A subset of the table is shown here. The complete table is
available electronically upon request.
\bigskip
\end{table}

\begin{table}[t]
\caption[ ]{Comparisons of our calculated WD formation rate with previous results}
\tabcolsep 20pt
\begin{tabular*}{140mm}{ccc}
\hline
Formation rate $(pc^{-3} yr^{-1})$     &   Sample  &   Reference   \\
\hline
\smallskip
$2.58\times 10^{-13}$ & 531 DA WDs from SDSS & this work\\
$(0.1 - 2)\times 10^{-10}$ &   89 WDs  &   Green (1980)    \\
$(3.9 - 6.1)\times 10^{-13}$   &   353 DA WDs from PG  &   Fleming et al. (1986)    \\
$2.3\times10^{-12}$ &   &   Weidemann (1991)    \\
$(7 - 10)\times 10^{-13}$  &   90 hot WDs from EUVE    &   Vennes et al. (1997)    \\
$6\times10^{-13}$   &   348 DA WDs from PG  &   Liebert et al. (2005)   \\
\hline
\end{tabular*}
\end{table}


\begin{thebibliography}{}
\bibitem[2003]{abazajian2003}
 Abazajian, K., Adelman-McCarthy, J., et al. 2003, ApJ, 126, 2081.
\bibitem[2005]{barstow2005} Barstow, M. A., Bond, H. E., Holberg, J. B.,
Burleigh, M. R., Hubneny, I., \& Koester, D. 2005, MNRAS, 362, 1134
\bibitem[1994]{barstow1994}
Barstow, M. A., Holberg, J. B., Fleming, T. A., Marsh, M. C.,
Koester, D., \& Wonnacott, D. 1994, MNRAS, 270, 499
\bibitem[2001]{bergeron2001}
Bergeron, P., Leggett, S. K., \& Ruiz, M. T. 2001, ApJS, 133, 413
\bibitem[1995]{blf}
Bergeron, P., Liebert, J., \& Fulbright, M. S. 1995a, ApJ, 444, 810
\bibitem[1992]{bsl}
Bergeron, P., Saffer, R. A., \& Liebert, J. 1992, ApJ, 394, 228
(BSL)
\bibitem[1995]{bergeron19951}
 Bergeron, P., Wesemael, F., \& Beauchamp, A. 1995b, PASP, 107,
1047
\bibitem[1990]{bergeron1990}
Bergeron, P., Wesemael, F., Fontaine, G., \& Liebert, J. 1990,
ApJ, 351, L21
\bibitem[2005]{boudreault}
Boudreault, S., \& Bergeron, P. 2005, 14th European Workshop on
White Dwarfs, ASP Conference Series, Vol. 334, 249
\bibitem[1995]{bragaglia1995}
Bragaglia, A., Renzini, A., \& Bergeron, P. 1995, ApJ, 443, 735
\bibitem[1935]{chandra1935}
Chandrasekhar, S. 1935, MNRAS, 95, 207
\bibitem[1939]{chandra1939}
Chandrasekhar, S. 1939, An Introduction
to the Study of Stellar Structure (Chicago: University of Chicago
Press). 
\bibitem[1993]{clemens1993}
Clemens, J. C. 1993, Ph.D. thesis, Univ. of Texas
\bibitem[2006]{eisenstein2006}
Eisenstein D. J., et al. 2006, ApJS, 167, 40
\bibitem[1997]{finley1997}
Finley, D. S., Koester, D., \& Basri, G. 1997, ApJ, 488, 375
\bibitem[1986]{fleming1986}
Fleming, T. A., Liebert, J., \& Green, R. F. 1986, ApJ, 308, 176
\bibitem[2001]{fontaine2001}
Fontaine, G., Brassard, P., \& Bergeron, P. 2001, PASP, 113, 409
\bibitem[1978]{gatewood1978}
Gatewood, G. D., \& Gatewood, C. V 1978, ApJ, 225, 191
\bibitem[1980]{green1980}
Green, R. F. 1980, ApJ, 238, 685
\bibitem[1961]{hs1961}
Hamada, T., \& Salpeter, E. E. 1961, ApJ, 134, 683
\bibitem[2004]{kleinman2004}
Kleinman, S. J., Harris, H. C., Eisenstein, D. J., Liebert, J., et
al. 2004, ApJ, 607, 426
\bibitem[1987]{koester1987}
Koester, D. 1987, ApJ, 322, 852
\bibitem[1979]{koester1979}
Koester, D., Schulz, H., \& Weidemann, V. 1979, A\&A, 76, 262
\bibitem[2001]{koester2001}
Koester, D., et al. 2001, A\&A, 378, 556
\bibitem[1980]{koester91}
Koester, D., \& Weidemann, V.  1991, AJ,
102, 1152
\bibitem[1975]{lamb1975}
Lamb, D. Q., \& van Horn, H. M. 1975, ApJ, 200, 306
\bibitem[2005]{liebert2005}
Liebert, J., Bergeron, P., \& Holberg, J. B. 2005, ApJS, 156, 47
\bibitem[2004]{madej2004}
Madej, J., Nalezyty, M., \& Althaus, L. G. 2004, A\&A, 419, L5
\bibitem[1997]{marsh1997a}
Marsh, M. C., Barstow, M. A., Buckley, D. A., Burleigh, M. R., et
al. 1997a, MNRAS, 286, 369
\bibitem[1997]{marsh1997b}
Marsh, M. C., Barstow, M. A., Buckley, D. A., Burleigh, M. R., et
al. 1997b, MNRAS, 286, 705
\bibitem[1995]{marsh1995}
Marsh, T. R., Dhillon, V. S., \& Duck, S. R. 1995, MNRAS, 275,
828
\bibitem[1999]{mccook1999}
 McCook, G. P., \& Sion, E. M. 1999, ApJS, 121, 1
\bibitem[1989]{mcmahan1989}
 McMahan, R. K.
1989, ApJ, 336, 409
\bibitem[1952]{mestel1952a}
Mestel, L. 1952, MNRAS, 112, 583
\bibitem[1999]{napiwotzki1999}
Napiwotzki, R., Green, P. J., \& Saffer, R. A. 1999, ApJ, 517,
399
\bibitem[2000]{panei20001}
 Panei, J. A.,
Althaus, L. G., \& Benvenuto, O. G. 2000, A\&A, 353, 970
\bibitem[2002]{phillips2002}
Phillips, J. P. 2002, ApJS, 139, 199
\bibitem[1994]{pottasch1996}
Pottasch, S. R. 1996, A\&A, 307, 561
\bibitem[1998]{provencal1998}
Provencal, J. L., Shipman, H. L., Hog, E., \& Thejll, P. 1998,
ApJ, 494, 759
\bibitem[1996]{reid1996}
Reid, I. N. 1996, AJ, 111, 5
\bibitem[1959]{schmidt1959}
Schmidt, M. 1959, ApJ, 129, 243 
\bibitem[1963]{schmidt1963}
Schmidt, M. 1963, ApJ, 137, 758
\bibitem[1968]{schmidt1968}
Schmidt, M. 1968, ApJ, 151, 393 
\bibitem[1975]{schmidt1975}
Schmidt, M. 1975, ApJ, 202, 22
\bibitem[1996]{schmidt1996}
Schmidt, H. 1996, A\&A, 311, 852
\bibitem[1997]{shipman1997}
Shipman, H. L., Provencal, J. L., Hog, E. \& Thejll, P. 1997, ApJ,
488, L43
\bibitem[1978]{thorstensen1978}
Thorstensen, J. R., Charles, P. A., Margon, B., \& Bowyer, S.
1978, ApJ, 233, 260
\bibitem[1999]{vennes1999}
Vennes, S. 1999, ApJ, 525, 995
\bibitem[1997]{vennes1997}
Vennes, S., Thejll, P. A., Galvan, R. G., \& Dupuis, J. 1997, ApJ,
480, 714
\bibitem[1999]{vennes1999}
Vennes, S., Thorstensen, J. R., \& Polomski, E. F. 1999, ApJ, 523,
386
\bibitem[1991]{vennes1991}
Vennes, S., Thorstensen, J. R., Thejll, P., \& Shipman, L. 1991,
ApJ, 372, L37
\bibitem[1980]{wegner1980}
Wegner, G.,  1980, AJ, 85, 1255
\bibitem[1989]{wegner1989}
Wegner, G.,  1989,
in IAU Colloq. 114, White Dwarfs, ed. G. Wegner (New Yor:
Springer), 401 
\bibitem[1987]{wegner1987}
Wegner, G., \& Reid, I.N. 1987, in IAU Colloq. 95,
The Second Conference on Faunt Blue Stars, ed. A. G. Davis
Phillip, D. S. Hayes, \& J. Liebert (Schenectady: L. Davis), 649
\bibitem[1991]{wegner1991} Wegner, G., \& Reid, I. N. 1991, ApJ, 375, 674
\bibitem[1989]{wegner1989}
Wegner, G., Reid, I. N., \& McMahan, R. K. 1989, 
in IAU Colloq. 114, White Dwarfs, ed.
G. Wegner (New Yor: Springer), 378
\bibitem[1989]{wegner1989}
Wegner, G., Reid, I. N., \& McMahan, R. K. 1991, ApJ, 376, 186
\bibitem[1990]{weidemann1990}
Weidemann, V. 1990, ARA\&A, 28, 103
\bibitem[1991]{weidemann1991}
Weidemann, V. 1991, in White Dwarfs, Proceedings of the 7th European 
Workshop, eds  G. Vauclair and E. Sion. NATO Advanced Science Institutes 
(ASI) Series C, 336, 67
\bibitem[1984]{weidemann1984}
Weidemann, V., \& Koester, D. 1984, A\&A, 132, 195
\bibitem[1990]{wood90}
Wood, M. A. 1990, Ph.D. Thesis, Texas Univ., Austin.
\bibitem[1995]{wood95}
Wood, M. A. 1995, Lecture Notes in Physics, 443,  41
\bibitem[2000]{york2000}
York, D. G., et al. 2000, ApJ, 120, 1579
\bibitem[2001]{zuckerman2001}
Zuckerman, B., \& Becklin, E. E. 2001, ApJ, 386, 260


\end{thebibliography}
\end{document}